\documentclass[11pt]{article}
\pdfoutput=1
\usepackage{jcapmod}
\usepackage{booktabs}
\usepackage{mathtools}
\usepackage[english]{babel}
\usepackage{amsmath,amssymb,amsbsy,amstext, amsthm,xcolor}
\usepackage{graphicx} 
\graphicspath{{plots/}}
\usepackage{amsfonts}
\usepackage{amssymb}
\usepackage{float}
\usepackage[utf8]{inputenc}
\RequirePackage{color}

\usepackage{colortbl}
\definecolor{green2}{cmyk}{0, 1, 0.5, 0}
\definecolor{lightgreen}{cmyk}{0.2, 0, 0.2, 0.2}
\definecolor{lightgray}{cmyk}{0.1,0.2,0,0.1}
\definecolor{lightgray2}{cmyk}{0.4,0.4,0,0.8}
\definecolor{black}{cmyk}{1.0,1.0,1.0,1.0}

\allowdisplaybreaks[1]

\usepackage{colortbl}
\definecolor{lightgreen}{cmyk}{0.2, 0, 0.2, 0.2}
\definecolor{lightgray}{cmyk}{0.1,0.2,0,0.1}
\definecolor{lightgray2}{cmyk}{0.1,0.1,0,0.1}

\makeatletter
\newlength{\apb@width}
\newcommand{\autoparbox}[2][c]{\settowidth{\apb@width}{#2}\parbox[#1]{\apb@width}{#2}}

\makeatother

\numberwithin{equation}{section}

\def\beq{\begin{equation}}
\def\eeq{\end{equation}}

\def\be{\begin{equation}}
\def\ee{\end{equation}}

\def\bea{\begin{eqnarray}}
\def\eea{\end{eqnarray}}
\def\eg{{\it e.g.~}}

\def\ie{{\it i.e.~}}
\def\d{{\rm d}}

\def\d{{\rm d}}

\def\nn{\nonumber}

\def\Mp{M_{\rm pl}}

\def\fr{\frac}

\def\0{{\boldsymbol 0}}

\def\fr{\frac}

\usepackage{setspace} 
\begin{document}

\begin{titlepage}

\setcounter{page}{1} \baselineskip=15.5pt \thispagestyle{empty}

\bigskip\

\vspace{1cm}
\begin{center}

{\fontsize{19}{28}\selectfont  \sffamily \bfseries {Squeezed tensor non-Gaussianity in non-attractor   inflation}}

\end{center}

\vspace{0.2cm}

\begin{center}
{\fontsize{13}{30}\selectfont Ogan \"Ozsoy$^{\star}$, Maria Mylova$^{\star}$, Susha  Parameswaran$^{\dagger}$, 
 Cari Powell$^{\star}$,  \\
Gianmassimo Tasinato$^{\star}$, Ivonne Zavala$^{\star}$} 
\end{center}

\begin{center}

\vskip 8pt
\textsl{$^\star$ Physics Department, Swansea University, Swansea, SA2 8PP, UK}\\
\textsl{$^\dagger$ Department of Mathematical Sciences, University of Liverpool, Liverpool, L69 7ZL, UK}
\vskip 7pt

\end{center}

\vspace{1.2cm}
\hrule \vspace{0.3cm}
\noindent
We investigate primordial tensor non-Gaussianity in  single field inflation,  during a phase of non-attractor evolution
 when  
the spectrum of primordial tensor modes can be enhanced
 to a level   detectable  at interferometer scales. Making use of a tensor 
duality we introduced in \cite{Mylova:2018yap}, we analytically compute the full bispectrum of primordial tensor fluctuations
during the non-attractor era. During this epoch  the shape of the tensor bispectrum is enhanced in the squeezed limit, its amplitude can be amplified
with respect to slow-roll   models, and tensor non-Gaussianity can exhibit a  scale dependence  distinctive of our set-up. 
We prove that our results do not depend on the frame used for the calculations.  Squeezed tensor non-Gaussianity
induces a characteristic quadrupolar anisotropy on the power spectrum of 
 the stochastic  background of primordial tensor perturbations. As a  step to make contact with gravitational wave experiments, we discuss  the 
 response function of a ground based Michelson interferometer to a  gravitational wave background with such a feature. 
\vskip 10pt
\hrule

\vspace{0.6cm}
 \end{titlepage}

 \tableofcontents

\newpage

\section{Introduction}

The possibility to directly detect a stochastic background of primordial 
tensor modes with  gravitational wave experiments  would offer new ways to
probe the physics of inflation. Such an opportunity would allow us to probe a much larger  range of
frequency scales than what  can be tested with CMB physics.
 Various scenarios have been proposed for enhancing the primordial  tensor
 spectrum at interferometer scales: from coupling the inflation to additional
 fields, whose dynamics are characterised by instabilities that amplify the tensor
 spectrum (see e.g. \cite{Cook:2011hg,Senatore:2011sp,Carney:2012pk,Biagetti:2013kwa,Biagetti:2014asa,Goolsby-Cole:2017hod,Sorbo:2011rz, Anber:2012du,Barnaby:2010vf,Barnaby:2012xt,Ozsoy:2017blg,Maleknejad:2011jw,Dimastrogiovanni:2012ew,Adshead:2013qp,Adshead:2013nka,Obata:2014loa,Maleknejad:2016qjz,Dimastrogiovanni:2016fuu,Agrawal:2017awz,Adshead:2017hnc,Caldwell:2017chz,Agrawal:2018mrg,Espinosa:2018eve}), to models
 that break space-time symmetries during inflation, leading to a blue spectrum
 for primordial tensor modes (see e.g. \cite{Endlich:2012pz,Bartolo:2015qvr,Ricciardone:2016lym,Ricciardone:2017kre,Domenech:2017kno,Ballesteros:2016gwc,Cannone:2015rra,Lin:2015cqa,Cannone:2014uqa,Akhshik:2014bla,Biagetti:2017viz,Dimastrogiovanni:2018uqy,Fujita:2018ehq}). See the general
 discussion in  \cite{Bartolo:2016ami}.

 In this work,  we focus on a  third possibility, and
  further develop on the idea introduced in \cite{Mylova:2018yap}. We  investigate single field 
 scenarios in which the inflationary expansion undergoes a brief phase of 
 non-attactor dynamics that amplify the tensor modes. Non-attractor cosmological evolution is
 known to enhance the  scalar sector of fluctuations, for example during ultra-slow roll
 or in constant roll inflationary systems \cite{Starobinsky:1992ts,Inoue:2001zt,Linde:2001ae,Kinney:2005vj,Martin:2012pe,Motohashi:2014ppa,Yi:2017mxs,Dimopoulos:2017ged,Pattison:2018bct}: this property  has been exploited in models producing
 primordial black holes in single field inflation
 (see 
 e.g. \cite{Garcia-Bellido:2017fdg,Sasaki:2018dmp,Carr:2016drx} for  reviews
         and  \cite{Saito:2008em,Garcia-Bellido:2017mdw,Germani:2017bcs,Motohashi:2017kbs,Ballesteros:2017fsr,Ezquiaga:2017fvi,Cicoli:2018asa,Ozsoy:2018flq,Biagetti:2018pjj} for   specific models). 
         In \cite{Mylova:2018yap} we showed that a similar
 behaviour can apply to the primordial  tensor sector, if we non-minimally couple the inflationary scalar 
 field with gravity during inflation. 
 During the non-attractor phase, the amplitude of the would be decaying tensor mode  becomes  amplified
  instead of suppressed at superhorizon scales (while the usual
  growing mode has constant amplitude), and the total tensor spectrum can be enhanced
     to a level detectable with gravitational wave experiments.
  Interestingly, there exists a `tensor duality' (which generalises
  to the tensor sector a similar duality for the scalar sector \cite{Wands:1998yp}) which 
    maps the evolution of tensor fluctuations
   during  the non-attractor phase to the dynamics  of tensor fluctuations in a slow-roll phase of expansion. 
   We use   the duality to  obtain an
   analytic control on the physics of  tensor modes during the phase of non-attractor evolution
    -- even if we are  far from a slow-roll approximation --
    and to {\it analytically compute} the properties of tensor non-Gaussianity during the non-attractor phase. 
    Tensor non-Gaussianity is an interesting observable which can help to characterise and
    distinguish different scenarios of inflation that enhance tensor modes at small or large frequency scales (see
    e.g. \cite{Maldacena:2002vr,Maldacena:2011nz,Soda:2011am,Shiraishi:2011st,Bartolo:2017szm,Agrawal:2017awz,Agrawal:2018mrg,Zhu:2013fja,Huang:2013epa,Cook:2013xea,Garcia-Bellido:2017aan,Bartolo:2018rku,Bartolo:2018evs}
    and 
    the review in Section 5 of \cite{Bartolo:2018qqn} for more a comprehensive reference list).  
In our framework,    tensor non-Gaussianity  is characterised by the following properties, which we are
going to discuss in what follows:
\begin{itemize}
\item The amplitude of the tensor bispectrum during the non-attractor evolution
 can be enhanced with respect slow-roll inflation, and its shape is amplified in a squeezed limit. 
 We analytically show that
 the   tensor bispectrum exhibits a characteristic scale dependence distinctive
 of our scenario, which can make our model quantitatively distinguishable
 from other frameworks with large tensor non-Gaussianity.
 \item  We show that  our results remain the same after applying 
a
 disformal plus a conformal transformation to our system.
 These transformations, at quadratic level in a perturbative expansion 
    in tensor fluctuations, render the system 
  identical to  Einstein gravity minimally coupled with a scalar field 
   \cite{Creminelli:2014wna,Baumann:2015xxa}. On the other hand, as we shall discuss,
 at {\it cubic} level in a perturbative expansion  tensor interactions include terms as $\dot{h}_{ij}^3$ which cannot be associated
 with contributions of standard Einstein gravity.
 \item The squeezed limit of the tensor bispectrum during non-attractor evolution
  does not satisfy Maldacena's consistency relations \cite{Maldacena:2002vr}, and can be parametrically
  amplified with respect to standard
  slow-roll scenarios. This is due to the fact that the would be decaying tensor
  mode is also excited in our system, and the corresponding dynamics is not a `single tensor'
  adiabatic system where Maldacena's arguments apply. This is analogous to what happens
  for models discussing  non-attractor inflation in the scalar sector.
  \item 
  Squeezed tensor non-Gaussianity
induces a characteristic quadrupolar anisotropy on the power spectrum of 
 the stochastic  background of primordial tensor perturbations. As a  step to make contact with gravitational wave experiments, we discuss  the 
 response function of a ground based Michelson interferometer to a  gravitational wave background with such a feature. 
\end{itemize}

\subsubsection*{\it Conventions}
We will use natural units, $\hbar=c=1$, with reduced Planck mass $\Mp^2 =(8\pi G)^{-1} \,=\,1$. Our metric signature is mostly plus  $(-,+,+,+)$. The background metric is a FRW universe with line element $d s^2\,=\,
-d t^2+a^2(t)\,d \vec x^2\,=\, a^2(\tau)\,\left( 
- d \tau^2+d \vec x^2
\right)
$. Throughout the paper, we adopt the following Fourier convention
\beq
q_{ij}({\bf x},t) = \int \fr{\d^3 k}{(2\pi)^3} ~ q_{ij}({\bf k},t)  ~ e^{-i{\bf k}.{\bf x}}.
\eeq

\section{Non-attractor dynamics and tensor fluctuations}\label{sec:review}

We discuss a new method \cite{Mylova:2018yap} for enhancing tensor fluctuations during inflation, which exploits
 the  structure of tensor  kinetic terms  in   inflationary theories
with non-minimal derivative couplings between scalars and gravity. 
 This Section is mainly intended as an enlightening    review of the methods and results developed
in \cite{Mylova:2018yap}. We  fix the notation and set the stage for the calculations
of tensor non-Gaussianity in  Section \ref{sec:nonG}.

\subsection{A mechanism for enhancing tensor fluctuations at super-horizon scales}

We 
focus on   spin-2 tensor fluctuations around a FRW cosmological background,  defined as \cite{Maldacena:2002vr}
\beq\label{met}
d s^2\,=\,-\d t^2+a^2(t)\,\left(e^{h} \right)_{ij} \,\d x^i \d x^j\,,
\eeq
with 
\be
\left(e^{h} \right)_{ij} \,=\,\delta_{ij}+h_{ij}+\frac12 h_{ik} h_{kj}+\frac16 h_{ik} h_{kl} h_{lj}+\dots\ee
where
 $h_{ij}$ is a transverse traceless spin-2 tensor perturbation. 
  At leading order in a derivative expansion, 
 the quadratic action for tensor perturbations can be expressed as  (see e.g. \cite{Kobayashi:2011nu}. From now on, 
 unless otherwise stated, we set $M_{\rm pl}=1$)
\bea
S_{T}&=&\fr{1}{8}\,\int \d t \,\d^3 x\,a^3(t)\,\,\left[ {\cal G}_T(t)\, \left( \partial_{t }
h_{ij} \right)^2-\frac{{\cal F}_T(t)}{a^2(t)}\left( \vec \nabla  h_{ij}  \right)^2\right]
\,,\nonumber \\
 \label{actssp}
&=&\fr{1}{2}\,\int \d y \,\d^3 x\,z_T^2(y)\,\left[  \left( \partial_{y }
h_{ij} \right)^2-\left( \vec \nabla  h_{ij}  \right)^2\right]\,,\nonumber
\\
& \underset{\left(z_T\,h_{ij}\,\equiv\,v_{ij} \right)}{=}&\,\,\frac12\,\int \d y \,\d^3 x
\,\left[  \left( 
v'_{ij} \right)^2- \,\left(\vec \nabla v_{ij}  \right)^2 +\frac{z_T''}{z_T}\,v_{ij}^2\right]\,.
\eea
The first line of this formula contains two functions of time ${\cal G}_T$, 
${\cal F}_T$ that characterise the tensor kinetic terms 
\footnote{The quadratic action  for tensor mode can be recasted into a canonical
`Einstein frame' action
 by
applying a disformal and a conformal transformation
to the system \cite{Creminelli:2014wna}. Nevertheless, 
all our results  remain the same in any frame one uses, as
 anticipated in \cite{Mylova:2018yap} and explained more at length
in Section \ref{sec:nonG} and Appendix \ref{app:disf} of the present paper.}. Their structure depends
on non-minimal couplings of gravity to the scalar field driving inflation, and on
the homogeneous scalar profile  (see Subsection \ref{sec:review-model} and \cite{Kobayashi:2011nu}).
In the  second line of eq  \eqref{actssp} we redefine the time variable as
 \bea \label{redtc}
 \d t&=&a\,\left(\frac{{\cal G}_T}{{\cal F}_T}\right)^{1/2}\,\d y
\,,
\\
&=&\frac{a(y)}{c_T(y)}\,\d y
 \eea
 in order to  express the action for tensor fluctuations as the one for a free field  in a time dependent background. We also  introduce  the convenient combination
 \beq \label{comz}
 z_T^2(y)\,=\,\frac{a^2(y)}{4}\,\sqrt{ {\cal G}_T(y)\,{\cal F}_T(y)} \,,
 \eeq
 dubbed as {\it tensor pump field} in analogy with the nomenclature used in the literature for the scalar
 sector.

Focussing on super-horizon evolution, defined in Fourier space as the condition
 $k^2\ll |z_T''/z_T|$,   the equations of motion for tensor
 fluctuations associated with action \eqref{actssp} admits the following solution
 \be \label{solfh}
 h_{ij}(y)\,=\,q_1+q_2\,\int^y\,\frac{\d y'}{z_T^2(y')}\,,
 \ee
 with $q_1$ and $q_2$ two integration constants, which can be fixed by matching
  with the solution at sub-horizon scales.
  $q_1$ corresponds to the usual constant
 mode at super-horizon scales, while the coefficient of $q_2$  would be the decaying mode,  if $z_T$ 
 were an increasing function of the time variable $y$. On the other hand, 
 whenever $z_T$ becomes a {\it decreasing} function of $y$, 
 we enter in a non-attractor phase where
 tensor modes  can grow on super-horizon scales:
 \be
 \frac{z_T'}{z_T}\,<\,0 \hskip1cm \Longrightarrow\hskip1cm {\text{Tensor modes grow on super-horizon scales\,.}}
 \ee 
 This is a {\it  non-attractor  phase} for tensor fluctuations, since the would be decaying mode actually
 increases and controls the amplitude of tensor fluctuations at large scales.  Such behaviour
 for the pump field $z_T$ 
 usually requires a departure from a slow-roll approximation, and the evolution and properties of fluctuations cannot
 be described in terms of usual slow-roll formulae. On the other hand, as shown in \cite{Mylova:2018yap}, we can make use
 of {\it tensor duality} -- which generalises to the tensor spectrum the scalar duality first pointed out
 in \cite{Wands:1998yp}  (see also \cite{Finelli:2001sr,Gasperini:2002bn,Gratton:2003pe,Boyle:2004gv,Piao:2004uq,Allen:2004vz,Khoury:2008wj,Khoury:2010gw}) -- to analytically investigate the dynamics of tensor modes in the non-attractor phase.
 
 Tensor duality is defined as follows. 
 In the 
 third line of eq \eqref{actssp} we rescale tensor modes as $h_{ij}\,\equiv\,v_{ij}/z_T$   in order to rewrite
 the action as a free system  in flat space with a time-dependent mass term for the mode $v_{ij}$. The mass parameter $z''_T/z_T$ depends
 on time, and its value can change during the inflationary evolution. We consider  two distinct phases of cosmological expansion, 
each lasting for a certain time interval,  
 characterised respectively by two regimes for the parameters $z_T$ and $\tilde{z}_T$ related by the {\it duality} condition
 \be
\frac{
 \tilde{z}_T''
 }{\tilde{z}_T}\,=\, \frac{
 {z}_T''
 }{{z}_T}
\hskip1cm \Longrightarrow 
\hskip1cm
\tilde{z}_T(y)\,=\,z_T(y) \left(
 c_1+c_2 \int^y\,\frac{\d y'}{z_T^2(y')}
 \right) \label{dual-cond}
 \ee 
 for  constant values of  $c_1$, $c_2$ (not to be confused with the $q_{1,\,2}$ of eq \eqref{solfh}). 
 Then the quantity  $v_{ij}$ is described by the very same action in the two phases (the third line
 of eq \eqref{actssp}) and the corresponding mode $h_{ij}$ is described by the same statistics in the two epochs -- only the
  time-dependent overall normalization  changes. The most useful application of such { tensor duality} is perhaps the tensor dual 
 of a slow-roll phase  characterised by constant functions ${\cal G}_T$ and ${\cal F}_T$, leading to the slow-roll  condition $ z_T\,=\,{\rm const}\, \times a(y)
 $.
   In the dual epoch, $|\tilde z_T|\,=\,{\rm const}/ a^2(y)$: we are in a non-attractor phase of evolution, since $\tilde z_T$ decreases with time. On the other hand, in both phases,
   \be \label{sec-der}
   \frac{\tilde z''_T}{\tilde z_T}\,=\, \frac{ z''_T}{ z_T}\,=\,\frac{2}{y^2}\,.
   \ee
    The corresponding spectrum of tensor modes $\tilde h_{ij}
 \,=\, \tilde z_T\,v_{ij}$ is  scale invariant (as in slow-roll),  and 
   its amplitude increases with time at super-horizon scales
    during the
   non-attractor phase
    \be  \label{powa6}
 {\cal P}_{\tilde h}(t) \,=\, 
 \left(
 \frac{a(t)}{a_0} \right)^6\,{\cal P}_h
 \ee
 with $a_0$ being the value of the scale factor at the onset of the non-attractor era, and ${\cal P}_h$
 is the (nearly constant) value of the tensor spectrum at superhorizon scales in the initial slow-roll phase.
 See  the technical Appendix \ref{app:tech}  and \cite{Mylova:2018yap}  for  details on the computation of the tensor power spectrum during the non-attractor phase, leading to the aforementioned amplification. 
 
\begin{figure}[h!]
\hskip0.1cm\includegraphics[width = 0.48 \textwidth]{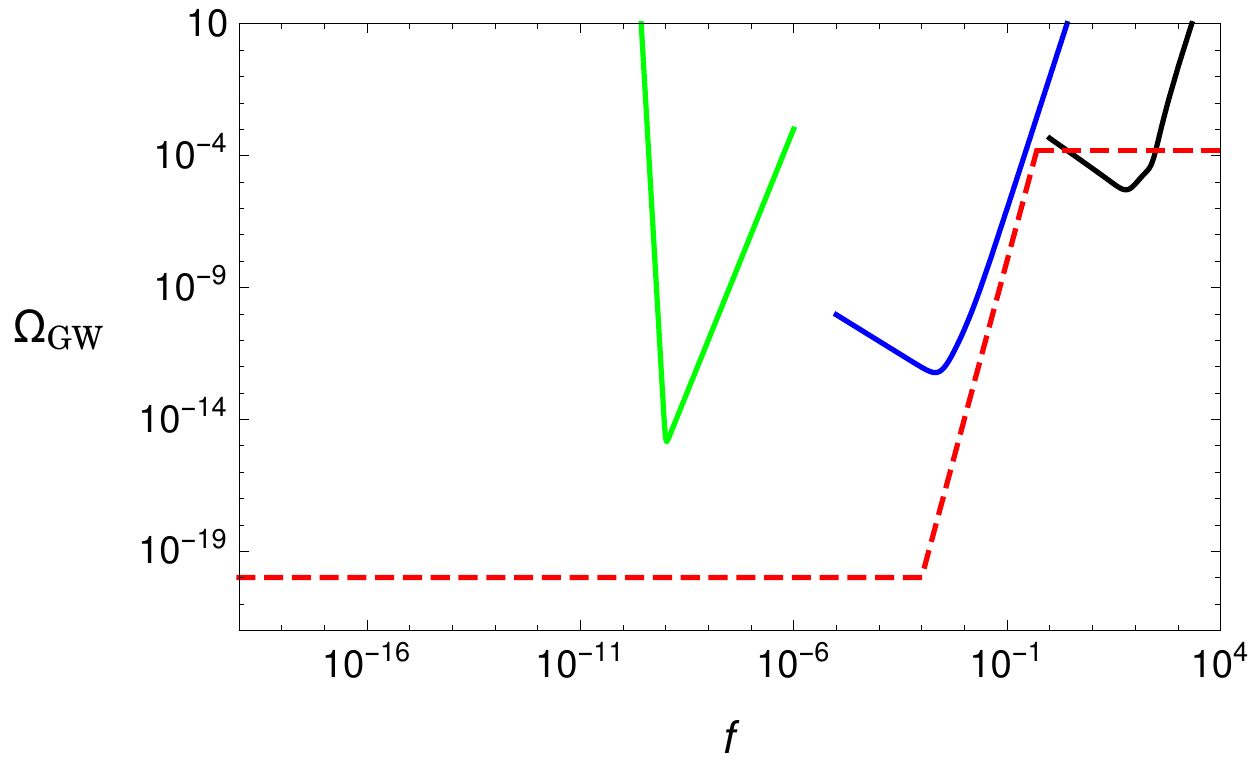}
\includegraphics[width = 0.48 \textwidth]{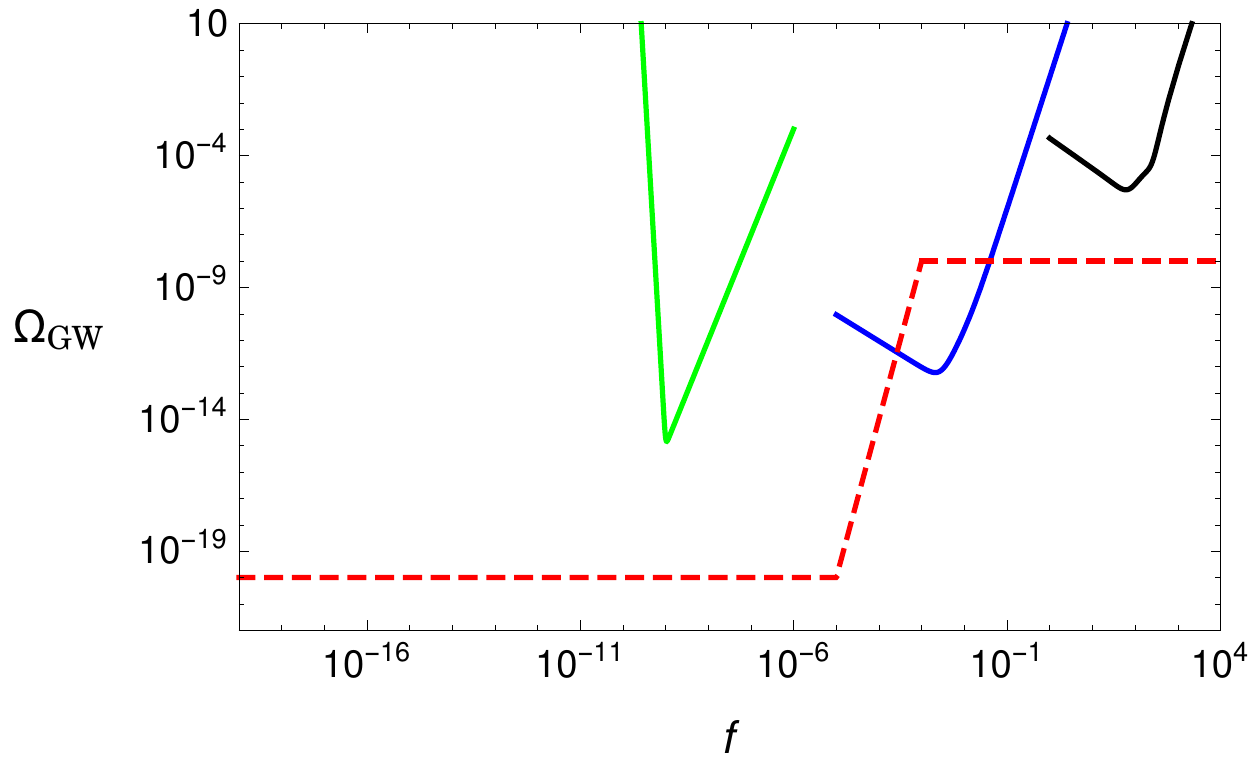}\\
\begin{center}
\hskip1cm\includegraphics[width = 0.70\textwidth]{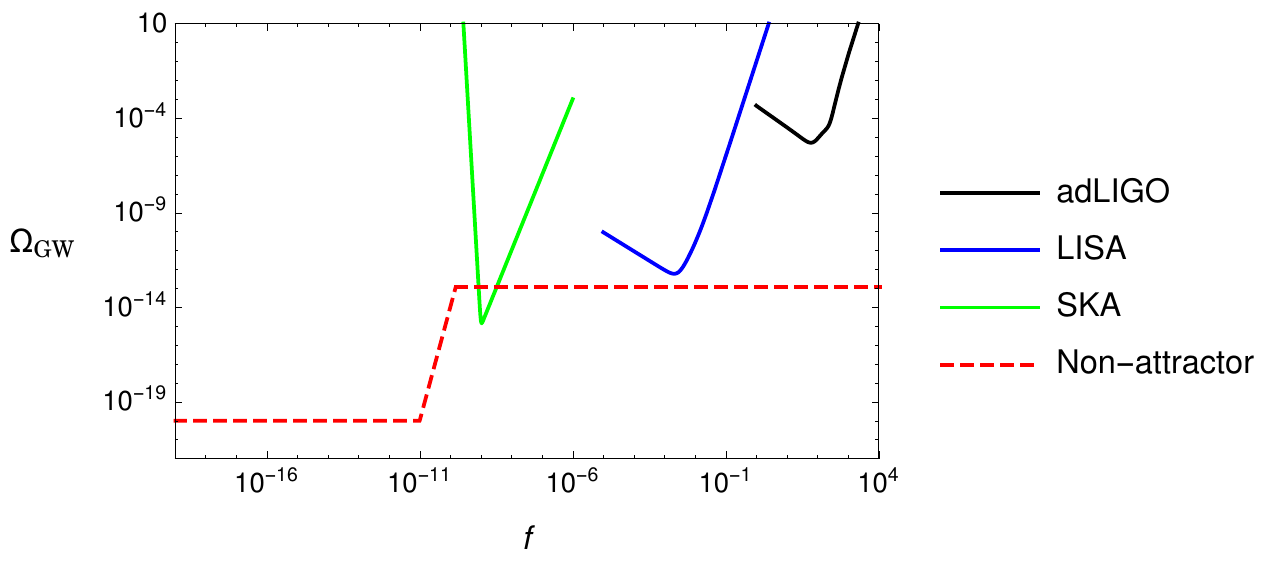}
 \caption{\it
 This figure shows qualitatively how 
  primordial tensor modes get amplified during a non-attractor phase. They 
  can contribute to the  GW energy density $\Omega_{GW}$, and thus enter within the sensitivity curves for 
GW detectors in their appropriate frequency ranges (expressed in Hz).  We model inflation as a pure de Sitter phase, during which a short period of non-attractor
evolution occurs -- whose starting time and duration depend on the model one considers -- enhancing the tensor spectrum. We use formula \eqref{powa6}, and assume for simplicity instantaneous transitions between attractor and non-attractor eras.  
  Our conventions for the definition of the GW energy density $\Omega_{GW}$
are the same as in \cite{Smith:2005mm}.}
\label{fig-enhanc}
\end{center} 
\end{figure}
 
 Such amplification of tensor fluctuations during a (typically short) non-attractor phase can lead to a very steep increase
 of the tensor spectrum as a function of frequency, once the amplitude of  primordial tensor modes is transferred to the present
 cosmological epoch  using standard formulae -- see e.g. \cite{Smith:2005mm}. 
 In Figure \ref{fig-enhanc}, we show
 how the spectrum of the superhorizon modes can be amplified to enter within the sensitivity curves for 
GW detectors, using formula \eqref{powa6}, and assuming for simplicity instantaneous transitions between attractor and non-attractor eras. 
The figure is only indicative, because it does not take into account the transition phases during different epochs and, above all, does
not take into consideration additional model-dependent constraints from amplification of scalar modes.   In 
 the next subsection we briefly  review an example of a concrete realisation of a tensor dual to a slow-roll phase
 in single field inflation.

 \subsection{A concrete realisation in single field inflation}\label{sec:review-model}
 
\begin{figure}[h!]
\begin{center}
\includegraphics[width = 0.7 \textwidth]{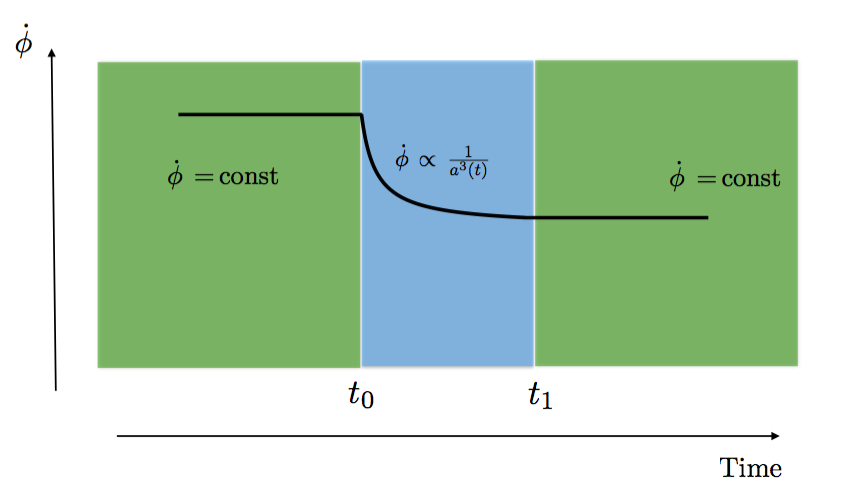}
 \caption{\it 
 Behaviour of the scalar field derivative in our system. 
 }
\label{fig:scal}
\end{center} 
\end{figure}

 We now briefly review an explicit  realisation of the mechanism of the previous subsection
 in a single field inflationary scenario, first presented in 
  \cite{Mylova:2018yap}. For convenience, we wish to find a single field inflationary model where the functions ${\cal F}_T$
 and ${\cal G}_T$ introduced in the action \eqref{actssp} are directly proportional to the square of scalar field velocity $\dot \phi$ 
 as
 \bea \label{fprowis}
 {\cal F}_{T}&\propto&\frac{\dot \phi^2}{H_0^2}\,,
 \hskip1cm;\hskip1cm
  {\cal G}_{T} \,\propto\,\frac{\dot \phi^2}{H_0^2}\,,
 \eea
 during the entire phase of the inflationary evolution, which for simplicity we describe in terms
 of pure de Sitter evolution with constant Hubble parameter $H_0$. The scalar field  
 follows a slow-roll evolution with constant velocity $\dot \phi\,=\,{\rm const}$ for most
 of the inflationary phase, but there is a brief phase of non-attractor evolution (whose duration is tunable in terms
 of the available parameters) during which $\dot \phi \propto 1/a^3$: 
 \bea
  \dot \phi\,\propto
  \left\{ 
\begin{array}{l} 
{\rm const}\hskip1cm {\text{during slow-roll phase}} \nonumber
\\
 1/a^3\hskip1cm {\text{during non-attractor phase}}  \nonumber
\end{array}
\right.
 \eea
  See Fig \ref{fig:scal} for a representation
 of the scalar field time dependent profile. Plugging these scalar profiles into the expressions for the functions
 ${\cal F}_{T}$, $ {\cal G}_{T}$ of eq \eqref{fprowis} and recalling the definition of the pump field $z_T$, eq 
 \eqref{comz}, it is easy to show that during the non-attractor phase we can use the tensor duality of eq \eqref{dual-cond}, and 
 the tensor power spectrum is enhanced by a factor $a^6(t)$ in this era \eqref{powa6}. Indeed, such a scenario
 is conceptually similar to the model of Starobinsky \cite{Starobinsky:1992ts}, designed to enhance scalar fluctuations
 during non-attractor inflation (see also \cite{Biagetti:2018pjj} and references therein).
 
 The conditions described above can be realised if the scalar field has non-minimal couplings with
 gravity during inflation. We consider a scenario based on   Horndeski theory  of gravity, the 
 most general scalar-tensor set-up with second order equations of motion, which 
 is
 described
 by the Lagrangian 
\bea
{\cal L}_{\text{tot}}&=&{\cal L}_2+{\cal L}_3+{\cal L}_4+{\cal L}_5 \label{LTOTh}
\,,
\\
{\cal L}_2&=&G_2\,
\,,
\\
{\cal L}_3&=&-G_3\,\Box \phi\,
\,,
\\
{\cal L}_4&=&G_{4}\,R+G_{4 X} \left[ \left( \Box \phi\right)^2 -\left( \nabla_\mu \nabla_\nu \phi \right)^2\right]
\,,
\\
{\cal L}_5&=&
G_5\,G_{\mu\nu}\,\nabla^\mu\nabla^\nu\,\phi-\frac{G_{5X}}{6}
\,
\left[
\left( \Box \phi \right)^3-3 \Box \phi\,\left(\nabla_\mu \nabla_\nu \phi \right)^2
+2 \left(\nabla_\mu \nabla_\nu \phi \right)^3
\right]
\,. \label{defL5}
\eea
The quantities $G_a\,=\,G_{a} (\phi,X)$ ($a=1,\dots 5$) are in principle arbitrary functions of the scalar field $\phi$ and 
\beq
 X\,=\,-\,\frac12 \,\partial_\mu \phi \partial^\mu \phi\,,
 \eeq
   $R$ is the Ricci tensor, $G_{\mu\nu}$ is the Einstein tensor, 
 and $G_{a\,X} = \partial G_a/\partial X$. For simplicity, in this work we focus on scenarios
 where a  shift symmetry  $\phi \to \phi + c$ is imposed, and the $G_i$ only depend on the  kinetic function $X$.
Inflation in scenarios  based on Horndeski
and Galileon Lagrangians have a long history,  starting from \cite{Burrage:2010cu} and the more general G-inflation \cite{Kobayashi:2010cm,Kobayashi:2011nu} models. Scenarios of ultra slow-roll, non-attractor G-inflation have been discussed in \cite{Hirano:2016gmv}.
  The quadratic Lagrangian for tensor fluctuations is described by our initial action \eqref{actssp}
 with 
 \bea \label{defFT}
 {\cal F}_T&=& 2\left[ G_4
 -X\,\ddot{\phi}\,G_{5X}
 \right]
 \\
  {\cal G}_T&=&
  2\left[ G_4
 -2 \,X\,G_{4X}
 -X\,H_0\,\dot{\phi}\,G_{5X}
 \right] \,.
 \eea

 In \cite{Mylova:2018yap} we showed that the conditions \eqref{fprowis} can be realised by choosing the following structure 
 for the functions  $G_a(X)$
   \bea\label{GF}
\nn G_2 &=&\rho X+\frac{\sqrt 2\,H_0^2\,\alpha}{3}\,\sqrt{X}
 -\nu  \,,\hskip1cm
\nn G_3\,=\,\frac{\sqrt{2}\,\delta}{3\,H_0}\,\sqrt{X} \,,
 \\
 G_4&=&-\frac{\beta}{6 H_0^2}\,X   \,,\hskip3.4cm
 G_5\,=\,\frac{\sigma}{\sqrt{2}\,H_0^3 }\,\sqrt{X}     \,,    
 \eea
 where the Greek letters are constant coefficients -- which can be different during the three different
 phases of evolution summarised in Figure \ref{fig:scal} -- and $H_0$ is the constant Hubble parameter during inflation.
 We refer the reader to \cite{Mylova:2018yap} for a detailed analysis of the system, with a quantitative discussion
 on the conditions necessary to avoid instability and to enhance the tensor power spectrum at superhorizon scales
 during the non-attractor phase. 
  Notice that besides
 the tensor modes, scalar modes are also typically enhanced in these scenarios,
 although with a smaller amplitude \cite{Mylova:2018yap}. 
 
\section{Tensor non-Gaussianity in  non-attractor inflation}\label{sec:nonG}

The mechanism we analysed in the previous Section shows that it is  possible to  
enhance the tensor power spectrum at  small (interferometer) scales by exploiting 
the behaviour of the would-be decaying mode, which can grow in a 
regime of non-attractor inflation. An interesting feature of our mechanism
is that there exists a {\it tensor duality} which allows us to obtain {analytical}
expressions for the tensor power spectrum even in regimes that are
far from a slow-roll period.  In this Section, we study the tensor bispectrum,
providing 
 {\it analytical expressions} for this quantity during the non-attractor phase,
and showing that the amplitude, shape  and scale dependence 
 of  the tensor bispectrum can
be different with respect to standard slow-roll inflation.  

\smallskip

The tensor bispectrum is an interesting theoretical quantity which
allows to  discriminate
between primordial and astrophysical sources of stochastic
gravitational wave backgrounds (SGWB) \cite{Bartolo:2018qqn}: if large tensor non-Gaussianity is detected, 
then it is likely that the SGWB has primordial origin, since an astrophysical  GW background  -- formed by
contributions from many unresolved sources -- is likely to be Gaussian from the central limit theorem. 
The properties of the tensor bispectrum -- shape, scale dependence, its value  in the squeezed limit -- 
are important for characterising the field content during inflation, and to further distinguish among different primordial sources that can amplify the tensor
spectrum at interferometer scales \cite{Bartolo:2018qqn}.

\smallskip

Remarkably, the cubic action for tensor fluctuations around FRW in single field inflationary theories
with second order equations of motion -- the starting
point for the computation of the bispectrum --  has a simple structure, and
contains only two contributions  \cite{Gao:2011vs,Gao:2012ib}

\begin{eqnarray}\label{cubic-action}
\nn S_T^{(3)}
&=& \int \d t \,\d^3 x\,a^3\,
\left[
\frac{\mathcal{ F}_T}{4a^2}\left(h_{ik}h_{jl}
-\frac{1}{2}h_{ij}h_{kl}\right)\partial_k\partial_lh_{ij}
+
\frac{\dot\phi XG_{5X} }{12} \dot h_{ij}\dot h_{jk}\dot h_{ki}
\right],
\\
&=&S_{T(\rm GR)}^{(3)}+
S_{T(\rm new)}^{(3)}\,.
\end{eqnarray}
 This action is obtained expanding the Horndeski Lagrangian density
\eqref{LTOTh} up to cubic order in fluctuations, and the functions  ${\cal F}_T$ and $G_5$ are given respectively in 
\eqref{defFT} and \eqref{defL5}. The result of standard single field inflation with canonical kinetic terms is obtained
setting ${\cal F}_T\,=\,1$ (recall that we choose units such that $M_{\rm pl}^2=1$).
 The structure of the first contribution, containing spatial derivatives only, is the same as the one obtained expanding the Ricci scalar at cubic order around a FRW background: 
this is the reason we denote it as  $S_{T(\rm GR)}^{(3)}$. The second contribution, $S_{T(\rm new)}^{(3)}$, is instead specific
to the Horndeski action: notice that it contains three time derivatives $\dot{h}_{ij}^3$, a feature to which we will return later.  
Tensor non-Gaussianity associated with the action \eqref{cubic-action} was studied in detail  in a slow-roll regime in 
\cite{Gao:2011vs,Gao:2012ib}, where it was found that the `GR' term gives a bispectrum enhanced
in the squeezed limit, while the `new' contribution gives a bispectrum peaked in equilateral configurations. In this work, instead, we will work out the tensor non-Gaussianity during a transient non-attractor phase, finding quite different
results.

\subsection{Amplitude  of tensor non-Gaussianity}\label{sec-ampng}

We discuss the amplitude and some properties
of tensor non-Gaussianity during an era of non-attractor inflation, which is dual to 
 a slow-roll phase as explained in Section \ref{sec:review}.  For simplicity, we focus on the case where
 the background geometry is exactly described by a de Sitter universe, with constant
 Hubble parameter $H_0$ (in \cite{Mylova:2018yap} we proved  that the equations of motion admit this solution
 for the scale factor).  
 We relegate all the technical details to Appendix \ref{app:tech}, and we discuss here the physical
consequences of our computation of the tensor three point function in Fourier space during a non-attractor
era. In order to express our results more concisely, 
 we define two polarisation modes as  (here, $e_{ij}^{(s)}$ is the polarization tensor with the helicity states $s=\pm $,
satisfying $e_{ii}^{(s)}(\mathbf{k})=0=k_je_{ij}^{(s)}(\mathbf{k})$.
See Appendix \ref{app:tech} for more information regarding our conventions on the polarisation tensors)
\beq
\xi^{(s)}({\bf k})\equiv h_{ij}({\bf k}) e^{*(s)}_{ij}({\bf k}),
\eeq



\noindent
which allow us to express the three point function  in the non-attractor era as
\bea
\langle \xi^{(s_1)}( \mathbf{k}_1)\xi^{(s_2)}( \mathbf{k}_2)\xi^{(s_3)}( \mathbf{k}_3)\rangle &=& (2\pi)^7 \delta({\bf k}_1+{\bf k}_2+{\bf k}_3)\fr{\mathcal{P}^{(\rm end)^2}_h}{\Pi_i~ k_i^3}~\bigg(\mathcal{A}^{s_1s_2s_3}_{(\rm new)}+\mathcal{A}^{s_1s_2s_3}_{(\rm GR)}\bigg)\,,\\
&\equiv&(2 \pi)^7\,\delta({\bf k}_1+{\bf k}_2+{\bf k}_3)\,B^{s_1s_2s_3}( \mathbf{k}_i)\,.
\label{DEFbis}
\eea
Hence we define the tensor bispectrum $B^{s_1s_2s_3}( \mathbf{k}_i)$
 as the coefficient of the $\delta$-function in the previous expression, which depends on the momenta as well as on the polarisation indices.

Using \eqref{TGR}, \eqref{TNEW}, the amplitudes $$\mathcal{A}^{s_1s_2s_3}_{(\rm new),(\rm GR)} \equiv  e^{*(s_1)}_{i_1j_1}(\mathbf{k}_1)e^{*(s_2)}_{i_2j_2}(\mathbf{k}_2)e^{*(s_3)}_{i_3j_3}(\mathbf{k}_3)\mathcal{A}^{(\rm new),(\rm GR)}_{i_1 j_i i_2 j_2 i_3 j_3}$$ can be calculated following the same methods of \cite{Gao:2012ib}.
 For our scenario, in the non-attractor regime, we find
\begin{align}\label{AB}
\nn \mathcal{A}^{s_1s_2s_3}_{(\rm new)}&= \mathcal{A}^{(\rm new)}(k_1,k_2,k_3) F(s_1k_1,s_2k_2,s_3k_3)\\
\mathcal{A}^{s_1s_2s_3}_{(\rm GR)} &= \mathcal{A}^{(\rm GR)}(k_1,k_2,k_3) \fr{(s_1k_1+s_2k_2+s_3k_3)^2}{2} F(s_1k_1,s_2k_2,s_3k_3), 
\end{align}
where
\beq
F(x,y,z)= \fr{1}{64}\fr{1}{x^2y^2z^2}(x+y+z)^3(x-y+z)(x+y-z)(x-y-z).
\eeq
   $\mathcal{A}^{(\rm GR)}$  and $\mathcal{A}^{(\rm new)}$  are obtained  in equations  \eqref{AGR} 
and \eqref{ANEW} respectively, which we rewrite here:
\begin{align}\label{AGR2}
\nn \mathcal{A}^{(\rm GR)}& = - \fr{K}{64}\Bigg[\left(1-\fr{3}{K^3}\sum_{i\neq j} k_i^2 k_j -\fr{6}{K^3}\Pi_i k_i\right) (-Ky_{\rm end})^2\\& ~~~~~~~~~~~-\fr{\pi}{4}\left(1-\fr{4}{K^3}\sum_{i\neq j} k_i^2 k_j -\fr{4}{K^3}\Pi_i k_i\right) (-Ky_{\rm end})^3+\dots\Bigg],
\end{align}
\begin{align}\label{ANEW2}
\nn\mathcal{A^{(\rm new)}}= -\fr{3H\mu^{(\rm end)}}{16\mathcal{G}^{(\rm end)}_T} &\Bigg\{\left(K^3-3\sum_{i\neq j} k_i^2 k_j -6~\Pi_i k_i\right)\\&-\fr{1}{4}\Bigg(K^3-5\sum_{i\neq j} k_i^2 k_j +\fr{2}{K^2}\sum_{i\neq j}k_i^3k_j^2\Bigg)(-K y_{\rm end})^2\Bigg\} .
\end{align}
In these formulae, $K\,=\,k_1+k_2+k_3$, 
$$\mu^{(\rm end)} \equiv\dot{\phi}^{(\rm end)}X^{(\rm end)}G_{5X}^{(\rm end)}$$
and the `end' indicates the end of the non-attractor phase: our results then quantify  the non-Gaussianity  accumulated
by the tensor modes during the non-attractor era. 
 Before proceeding, some observations are in order:
\begin{itemize}
\item The squeezed limit of the bispectrum does not satisfy Maldacena's consistency relations \cite{Maldacena:2002vr}. Indeed, computing
the bispectrum of eq \eqref{DEFbis} for $s_1=s_2$ in the limit of squeezed isosceles triangles,  
we find 
\bea
B^{s_1s_1s_3}({\bf k}_1,\,-{\bf k}_1,\,{\bf k}_3\to 0 )&=&
\,\frac{
\left(
{\cal P}_{h}^{({\rm end})}
\right)^2
}{32\,k_1^3\,k_3^3}\,
\left[
\fr{3\,H \mu^{(\rm end)}}{\mathcal{G}_T^{(\rm end)}}
\left( 1+\frac{(-k_1 y_{\rm end})^2}{2}
\right)
+\frac{(-k_1 y_{\rm end})^2}{2}
\right]\,,\nonumber\\
\label{sqxbisf}
\eea
while we find zero for $s_1\neq s_2$. Instead, Maldacena's consistency relation  (with our conventions)
 would read in this case
  $$B^{s_1s_1s_3}({\bf k}_1,\,-{\bf k}_1,\,{\bf k}_3\to 0 )\,=\,\frac{3}{64\,k_1^3\,k_3^3}\,
\left(
{\cal P}_{h}^{({\rm end})}
\right)^2
\,.$$
 This can be expected, since during the non-attractor era, besides 
the usual growing tensor mode,  the would be tensor decaying mode
is excited as well, and we are no longer working in a `single tensor'   adiabatic system
 where Maldacena's arguments apply \footnote{Similar considerations have
 been developed   in various works for the
 scalar sector, see e.g. \cite{Namjoo:2012aa,Chen:2013aj,Martin:2012pe,Huang:2013lda,Chen:2013eea}, finding non-attractor models  with an enhanced  scalar bispectrum 
 in the squeezed limit.}. 
 By tuning the parameters of the model, this implies that the amplitude of the tensor
 bispectrum can be enhanced in the squeezed limit (see also  
 \cite{Ricciardone:2016lym,Ricciardone:2017kre,Dimastrogiovanni:2018gkl,Goon:2018fyu,Anninos:2019nib} for different
 scenarios with enhanced squeezed tensor bispectrum), with potentially interesting
 phenomenological consequences that we shall  discuss in Section \ref{sec:response}. 
\item While  the scenario studied so far is characterised by non-standard kinetic terms for the tensor
sector, it is  known that by performing a conformal followed
by a disformal transformation the {\it second order
action} for tensor modes -- our eq. \eqref{actssp} -- acquires the very same structure of the second
order action  of Einstein gravity around FRW homogeneous backgrounds \cite{Creminelli:2014wna}.   On the other hand,   the {\it third order action} we are considering here, eq \eqref{cubic-action},   contains a piece with three time derivatives $\dot{h}_{ij}^3$ -- an  
operator that cannot be recast into a pure `GR' contribution via disformal/conformal transformations.  This said, in Appendix \ref{app:disf} we show in detail that all our results remain the same also in an `Einstein frame' with standard 
second order tensor action: the only difference is that in this frame the non-attractor phase corresponds
to a short period of universe contraction.  
 \item  Our expressions for the tensor bispectrum contain a characteristic scale dependence with overall factors containing powers
 of $(-K \,y_{\rm end})$, that are distinctive of our scenario -- being absent in other frameworks with large tensor
 non-Gaussianity. The explicit dependence on the time $y_{end}$ when the non-attractor
 phase ends is due to the fact that the bispectrum has been computed specifically
 {at the end of}  the non-attractor phase.  For simplicity, we assume that this era is immediately followed by a standard slow-roll inflation, where tensor modes and
 their statistics are frozen in a super-horizon regime. The overall scale dependence 
 of the tensor bispectrum controlled by $K$ is also
 distinctive of our set-up.
 Similar   situations have been encountered in the scalar sector, starting 
 from the work  \cite{Khoury:2008wj}, for models with non-standard
 cosmological expansion history,  leading to interesting observables associated with scale-dependent
 non-Gaussianity (explored in general
 terms in \cite{Chen:2005fe,Byrnes:2009pe,Byrnes:2010ft}). A consequence of this fact is that non-Gaussianity depends on the scale and might
 be different at different interferometer  scales (for example, LIGO-VIRGO and LISA): it 
 would be interesting to further explore phenomenological
 consequences of this property, which goes beyond the scope of our work. 
\end{itemize}

We now continue by estimating  the amplitude of non-Gaussian signal.
As a measure for the amount of non-Gaussianity, we  use the following definition of the non-linearity parameter $f_{\rm NL}$, as  in the works \cite{Gao:2011vs,Gao:2012ib}
\beq\label{fnlg}
\hat{f}^{s_1s_2s_3}_{\rm NL(\rm new),(\rm GR)} \equiv 30 \fr{\mathcal{A}^{s_1s_2s_3}_{(\rm new),(\rm GR)}}{K^3}\bigg \rvert_{k_1=k_2=k_3},
\eeq
which is analogous to the standard $f_{\rm NL}$ for the scalar curvature perturbation. Notice that the 
$f_{\rm NL}$ above is defined in terms of equilateral configurations for tensor bispectra, and its value
depends on the polarisations. 
 Using the definitions of the two amplitudes in \eqref{AB}, we find 
\begin{align}
\hat{f}^{s_1s_2s_3}_{\rm NL(\rm new)}&=-\fr{45}{32}~  \big[3+ 2(s_1s_2+s_1s_3+s_2s_3)\big]\fr{\mathcal{A}^{(\rm new)}(K/3,K/3,K/3)}{K^3},\\
\hat{f}^{s_1s_2s_3}_{\rm NL(\rm GR)}&=-\fr{5}{64} \big[21+ 20(s_1s_2+s_1s_3+s_2s_3)\big] \fr{\mathcal{A}^{(\rm GR)}(K/3,K/3,K/3)}{K}.
\end{align}
The dependence of the non-linerity parameter on the polarization implies the following symmetry $\hat{f}^{++-}_{\rm NL(\rm new),(\rm GR)}=\hat{f}^{+--}_{\rm NL(\rm new),(\rm GR)}$ and   $\hat{f}^{+++}_{\rm NL(\rm new),(\rm GR)}=\hat{f}^{---}_{\rm NL(\rm new),(\rm GR)}$ which follows from the fact that the interactions we consider do not violate parity. More concretely, we have
\beq
\hat{f}^{+++}_{\rm NL(\rm new)} = \fr{135}{512}
\fr{H \mu^{(\rm end)}}{\mathcal{G}_T^{(\rm end)}}
\left(1 + \fr{5}{36}(-Ky_{\rm end})^2\right)
\eeq
and 
\beq
\hat{f}^{++-}_{\rm NL(\rm new)} = \fr{15}{512}\fr{H \mu^{(\rm end)}}{\mathcal{G}_T^{(\rm end)}}\left(1 + \fr{5}{36}(-Ky_{\rm end})^2\right).
\eeq
Similarly for $f^{s_1s_2s_3}_{\rm NL(\rm GR)}$ we have
\beq
\hat{f}^{+++}_{\rm NL(\rm GR)} = \fr{45}{4096}~(-Ky_{\rm end})^2~ \bigg[1 + \fr{\pi}{12}(-Ky_{\rm end})\bigg]
\eeq
and
\beq
\hat{f}^{++-}_{\rm NL(\rm GR)} = \fr{5}{36864}~(-Ky_{\rm end})^2~ \bigg[1 + \fr{\pi}{12}(-Ky_{\rm end})\bigg].
\eeq
These results show that the $f_{\rm NL}$ parameter is generically positive during the non-attractor phase, similar to the case of contracting universes considered in \cite{Khoury:2008wj}. Importantly, due to the strong scale dependence of the $f_{\rm NL(\rm GR)}$, $f_{\rm NL(\rm new)}$ dominates the bispectrum for $-k_i y_{\rm end}\ll 1$ and $H \mu^{(\rm end)}/\mathcal{G}_T^{(\rm end)}\sim \mathcal{O}(1)$. Recall that $\mathcal{G}_T \supset \mu H$, in particular
\beq\label{gt}
\mathcal{G}_T = 2\left( G_4 - 2XG_{4X}-\mu H \right)
\eeq
for the background model we discussed earlier. The expression above \eqref{gt}  indicates that  we need accidental cancellations\footnote{Note that this situation is not special to the model under consideration in this work and arises for general slow-roll scenarios as well \cite{Gao:2011vs}.} between the first two terms in $\mathcal{G}_T>0$ and $\mu H$ in order to ensure $H\mu/\mathcal{G}_T \gg 1$. We discuss in Appendix \ref{app:concrete} a concrete
scenario leading to  large tensor non-Gaussianity within the framework we reviewed in Section \ref{sec:review-model}.


\begin{figure}[t!]
\begin{center}
\includegraphics[width = 0.56 \textwidth]{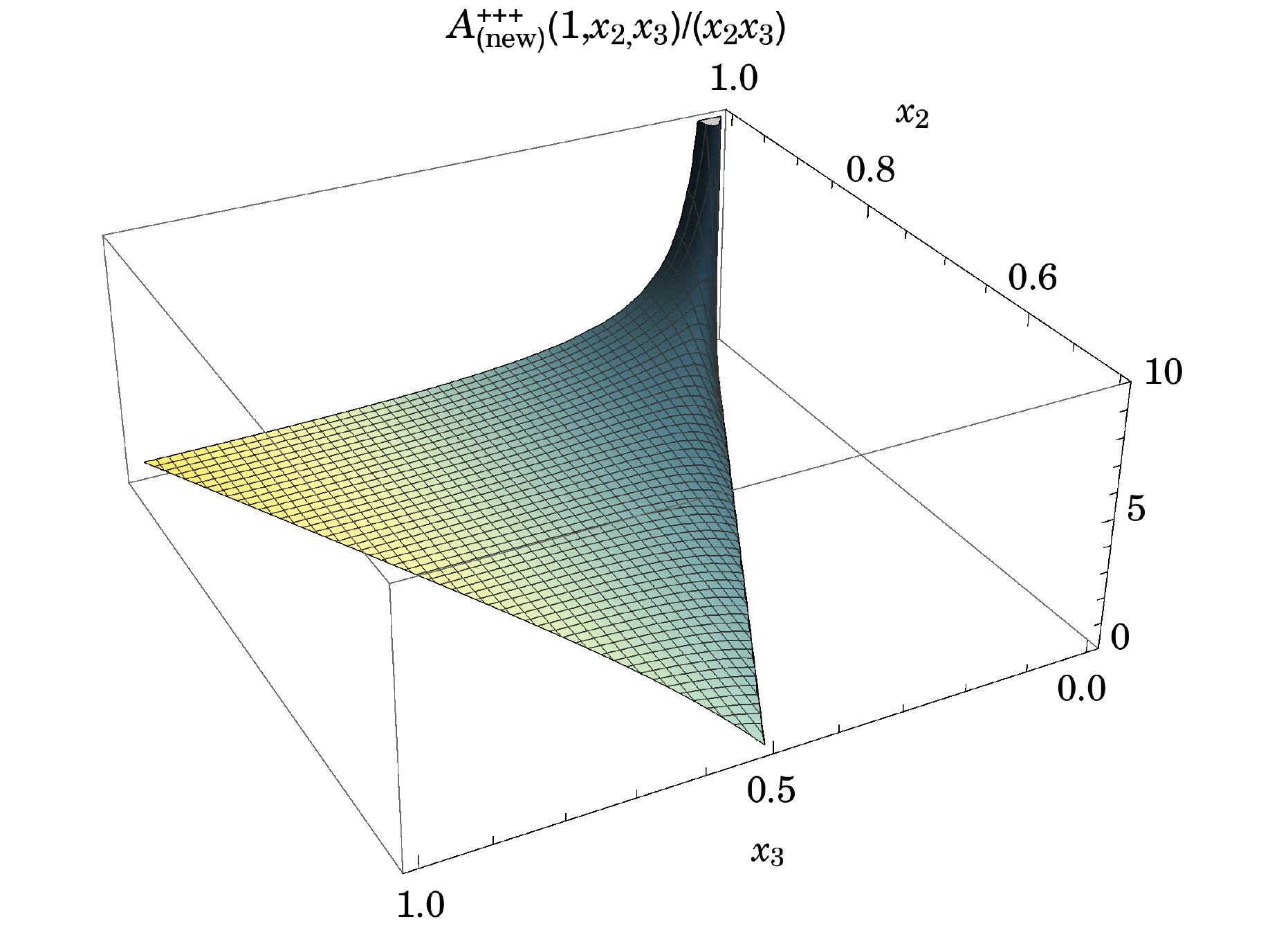}~
\includegraphics[width = 0.5 \textwidth]{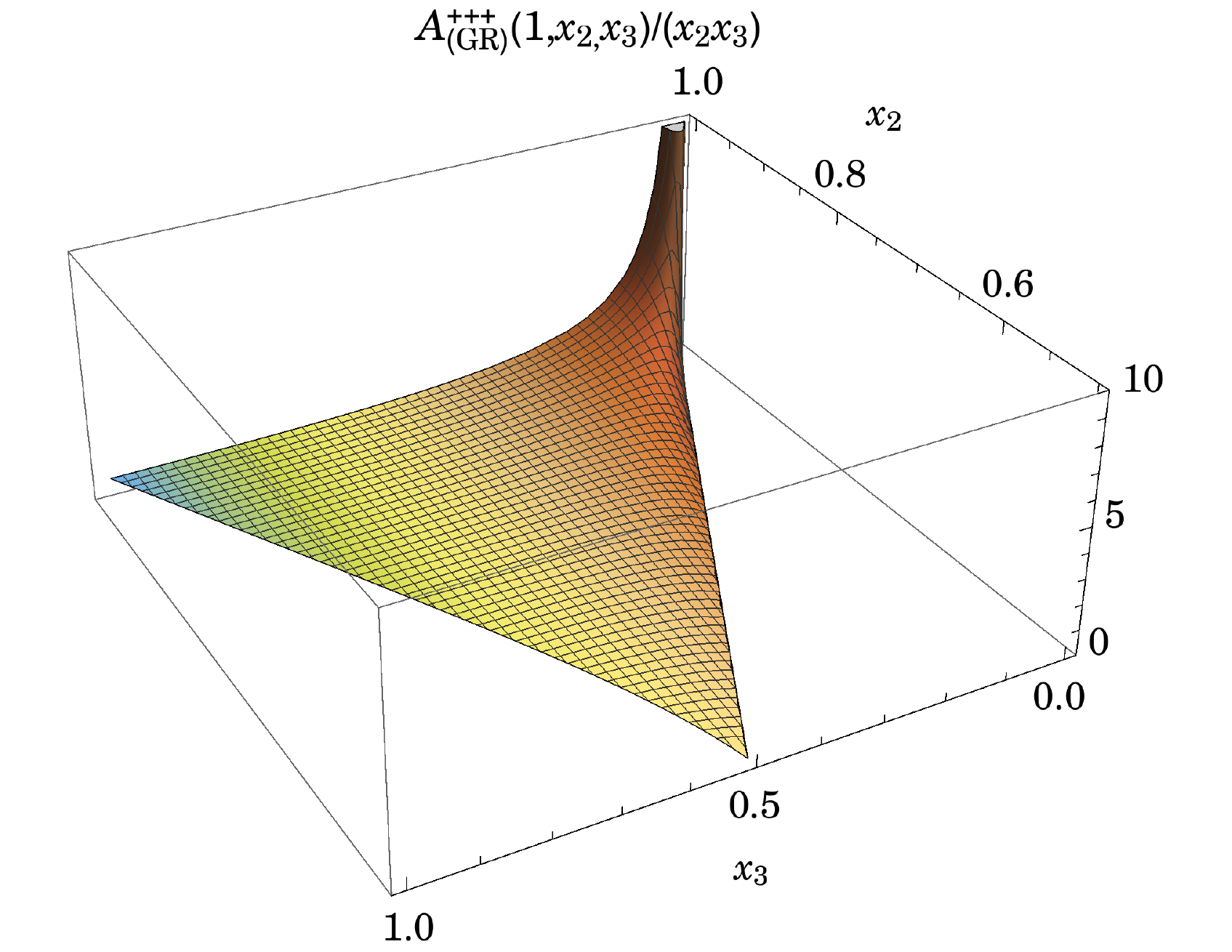}
\end{center}
\caption{\it $\mathcal{A}^{+++}_{(\rm new)}(1,x_2,x_3)/(x_2x_3)$ as a function of $x_2$ and $x_3$ where we set $H\mu^{\rm end}/\mathcal{G}_T^{\rm end}\to 1$ (Left). $\mathcal{A}^{+++}_{(\rm GR)}(1,x_2,x_3)/(x_2x_3)$ as a function of $x_2$ and $x_3$ (Right). In both plots we normalized the amplitudes $\mathcal{A}^{+++}_{(\rm new)}$ and $\mathcal{A}^{+++}_{(\rm GR)}$ to unity for equilateral configurations $x_2=x_3=1$ and took $-Ky_{\rm end}=10^{-2}$.\label{fig:NG}}
\end{figure}

\subsection{Shape of tensor non-Gaussianity}

We now study the shape of non-Gaussianity in our model. Since both of the amplitudes have  non-trivial scale dependence, we examine the shape of the amplitudes~\footnote{Recall that we are interested in modes that leave the sound-horizon during the non-attractor era, \ie $-k_i y_0 >1$ and $-k_i y_{\rm end} <1$. This implies that the modes of interest satisfy $1 > - K y_{\rm end} > e^{-\Delta N}$, where $\Delta N$ is the duration of the non-attractor era.} at fixed $-K y_{\rm end}$. We focus on the dimensionless ratio $\mathcal{A}^{s_1s_2s_3}_{(\rm new),(\rm GR)}/(k_1k_2k_3)$ of both amplitudes in 
\eqref{AB} following the literature for scalar perturbations \cite{Babich:2004gb}. In particular, we will plot $\mathcal{A}^{s_1s_2s_3}_{(\rm new),(\rm GR)}/(k_1k_2k_3)$ in the $x_2-x_3$ plane where $x_j \equiv k_j/k_1$ with $j\neq 1$ by restricting ourselves to the range $1-x_2\leq x_3 \leq x_2$. Note that the first inequality follows from the triangle inequality whereas the latter allows us to avoid plotting the same configuration twice. The non-Gaussian amplitudes $\mathcal{A}^{+++}_{(\rm new)}/(k_1k_2k_3)$ and $\mathcal{A}^{+++}_{(\rm GR)}/(k_1k_2k_3)$ are shown in Figure \ref{fig:NG}. We see that  both the interaction terms in \eqref{cubic-action} give rise to non-Gaussianity that peaks in the squeezed limit. This result is in contrast with the slow-roll case where the new contribution peaks in the equilateral configuration. This difference is due to the fact that during the non-attractor phase, the fluctuations in $h_{ij}$ keeps growing outside the horizon
due to the dynamics of the would be decaying mode,
  \ie $\dot{h}_{ij} = 3 H h_{ij}$ and therefore the non-Gaussian amplitude peaks when one of the wave numbers is small, corresponding to the squeezed-triangle limit\footnote{See \cite{Khoury:2008wj} for similar dynamics that lead an to enhanced squeezed bispectrum in curvature perturbations.}. In the standard attrac\color[rgb]{0,0,0}tor slow-roll background however, tensor fluctuations freeze on large scales, $\dot{h}_{ij}\to 0$ and therefore only wave-numbers comparable to the size of the horizon can contribute to the non-Gaussianity for the interaction proportional to the time derivatives of $h_{ij}$ in \eqref{cubic-action}. 

\subsection{Interferometer response function for  anisotropic tensor power spectrum}
\label{sec:response}

As we explained in Subsection \ref{sec-ampng}, our system does not satisfy Maldacena consistency relations:
the squeezed limit of the bispectrum can be enhanced by the contributions  of the would be decaying tensor mode. 
 This means that we can develop a scenario where at the same time we have a large tensor power spectrum
 at interferometer scales, accompanied by 
  enhanced  squeezed tensor non-Gaussianity. 
   In this subsection, we start with a  brief `theory' part to connect the squeezed limit of the tensor bispectrum
   with a quadrupolar anisotropy of the tensor power spectrum;   we then continue 
   with a discussion on  possible ways to detect an anisotropic  gravitational
   wave power spectrum with ground based interferometers, building
   on the results of \cite{Allen:1996gp}.
  
  \smallskip
  \noindent
  {\bf Theory:}
  A large non-Gaussianity in the squeezed limit can induce couplings
  between modes at different scales: the tensor power spectrum is  modulated  by
  long tensor modes that induce large scale anisotropies.
  This fact has been explored in several contexts, mainly in the scalar, but also
  in the tensor sector: see e.g. \cite{Giddings:2010nc,Gerstenlauer:2011ti,Dai:2013kra,Dimastrogiovanni:2014ina}.  
  Other scenarios that can induce large tensor non-Gaussianity in the 
  squeezed limit, by violating the adiabaticity condition in the tensor sector, are supersolid
  inflation \cite{Endlich:2012pz,Bartolo:2015qvr,Ricciardone:2016lym,Ricciardone:2017kre,Cannone:2014uqa,Lin:2015cqa}, bigravity or higher spin theories \cite{Biagetti:2017viz,Dimastrogiovanni:2018uqy,Dimastrogiovanni:2018gkl,Fujita:2018ehq,Goon:2018fyu,Anninos:2019nib}; our considerations can apply to these cases as well.

  When focussing on the `GR' operator of action, one finds that squeezed non-Gaussianity
  induces a quadrupolar anisotropy in the tensor power spectrum, with\footnote{Here $P_h({\bf k})\equiv \frac{2\pi^2{\cal P}_h({\bf k})}{\bf k^3}$ and $h_{\bf q}=v_{\bf q}a$ (see Appendix \ref{app:tech}).}   (see e.g.
  \cite{Bordin:2016ruc,Dimastrogiovanni:2018uqy})
 \be\label{tens-anis}
 {\cal P}_{h}({\bf{k}},\,{\bf x})
 \,=\, {\cal P}_{h}^{(0)}({{k}})\, \left[1+{\cal Q}_{ij}({\bf{k}},\,{\bf x}) \,\hat k_i \hat k_j \right]
 \ee
 and
 \be
 {\cal Q}_{ij} ({\bf{k}},\,{\bf x} )\,=\,\sum_{s_1,s_2} \int_{L^{-1}}\,\frac{d^3 q}{(2\pi)^3}\,e^{i\,{\bf q}
 \,{\bf x}}\,
 \left(
 \frac{B^{s_1 s_1 s_2}({\bf q},\,{\bf k},\,-{\bf q}-{\bf k})}{P_h(q) P_{h} (k)}
 \right)
 \,h_{\bf q} e_{ij}^{(s_2)} ({\bf q})\,.
 \ee
 We expect that the modulation \eqref{tens-anis} of the tensor power spectrum arises in any scenario with enhanced 
 squeezed tensor non-Gaussianity. 
    The  integral defining the anisotropy parameter  
     is evaluated in  a patch  centered at the position ${\bf x}$ and 
    spans over   long tensor modes with momenta within the non-attractor phase, corresponding to scales   well  larger than   
 the gravitational wave wavelengths
  under consideration (see e.g. Section 4.4 of \cite{Dimastrogiovanni:2018uqy}).   Being dependent on
   a linear combination of 
   the polarization tensors $e_{ij}^{(s)}$, the quantity $Q_{ij}$ is traceless.  It is convenient
 to define the squeezed limit of the bispectrum as
  \be
\lim_{{\bf q}\to0}  \frac{B^{s_1 s_1 s_2}({\bf k},\,-{\bf k}, {\bf q})}{P_h(q) P_{h} (k)}\,=\,
\frac{3}{64}
+\hat f_{\rm NL}^{\rm sqz}\,,
\ee
where the quantity $\hat f_{\rm NL}^{\rm sqz}$ parameterises the deviation from the Maldacena's consistency 
conditions. In our case, the quantity $\hat f_{\rm NL}^{\rm sqz}$  
can be read from eq \eqref{sqxbisf}.
The anisotropy parameter ${\cal Q}_{ij}$ is determined in a statistical sense,
 averaging over many large patches. Its  average equal to zero, and its  variance  results (see e.g.
 \cite{Bordin:2016ruc,Dimastrogiovanni:2018uqy})
\bea
\langle {\cal Q}_{ij}({ \bf k })\,{\cal Q}_{mn}({ \bf k }) \rangle
&=&
\frac{\pi}{20} \left( 
 \delta_{im} \delta_{jn}+\delta_{in} \delta_{jm}
-\frac23 \delta_{ij} \delta_{mn}
\right)
\,\int \frac{d k}{k}\,\big| \hat f_{\rm NL}^{\rm sqz} \big|^2\,{\cal P}_h^2\,,
\\ 
&\simeq&  \left( 
 \delta_{im} \delta_{jn}+\delta_{in} \delta_{jm}
-\frac23 \delta_{ij} \delta_{mn}
\right) \big| \hat f_{\rm NL}^{\rm sqz} \big|^2\,{\cal P}_h^2\, \Delta N
\eea
where in the last line we specialised for simplicity to the case of scale invariant power-spectrum and squeezed $f_{\rm NL}^{\rm sqz}$,
and $\Delta N$ indicates the number of e-folds of cosmological evolution associated with the non-attractor era. 
The value of ${\cal P}_h$ in the previous formula indicates the magnitude of the tensor power spectrum at the
end of non-attractor, which can be much larger than its value during the initial phase of slow-roll. Assuming that the  magnitude of tensor spectrum is of order $\sim10^{-14}$ at large scales, and it receives a $10^{10}$ enhancement 
during three e-folds of non-attractor inflation (using eq \eqref{powa6}), we learn that $f_{\rm NL}^{\rm sqz}\sim100$
is sufficient to give a value for $\sqrt {\langle  {\cal Q}_{ij}^2 \rangle}$ of the order of a few percent (but $\sqrt {\langle  {\cal Q}_{ij}^2 \rangle}$ can be larger
depending on the magnitude of tensor non-Gaussianity).

These results   imply that the size of the anisotropy parameter can be a probe of the squeezed tensor bispectrum. We now
outline  a possible way to test   such quantity with ground based interferometers
\footnote{Tensor non-Gaussianity  can  also be an important observable for characterizing the primordial 
stochastic gravitational wave background at CMB scales, and have been explored in other contexts, see e.g.  \cite{Thorne:2017jft,Agrawal:2017awz,Agrawal:2018mrg}.}.

  \smallskip
  \noindent
  {\bf Connection with gravitational wave experiments:}
The possibility of detecting anisotropies in   a
 SGWB has started with the work
  \cite{Allen:1996gp}, that derived
  the formalism necessary to quantitatively address the subject. The  motivation for
  such investigations  is  to detect signals  from  a stochastic background due to   astrophysical sources that can
  generate multipolar anisotropies. On the other hand, the  
  formalism of  \cite{Allen:1996gp} is sufficiently general and  can be used with little changes  also to investigate  
   tensor anisotropies from the early universe. We apply the formulae and arguments of  \cite{Allen:1996gp}  to analyse tensor
  power spectra with a quadrupolar
  anisotropic structure as in eq \eqref{tens-anis}. We focus for simplicity on
analysing the response function for a single 
 Michelson
ground-based interferometer  (see \cite{Maggiore:1999vm,Maggiore:1900zz} for reviews).

\begin{figure}[h!]
\begin{center}
\includegraphics[width = 0.5 \textwidth]{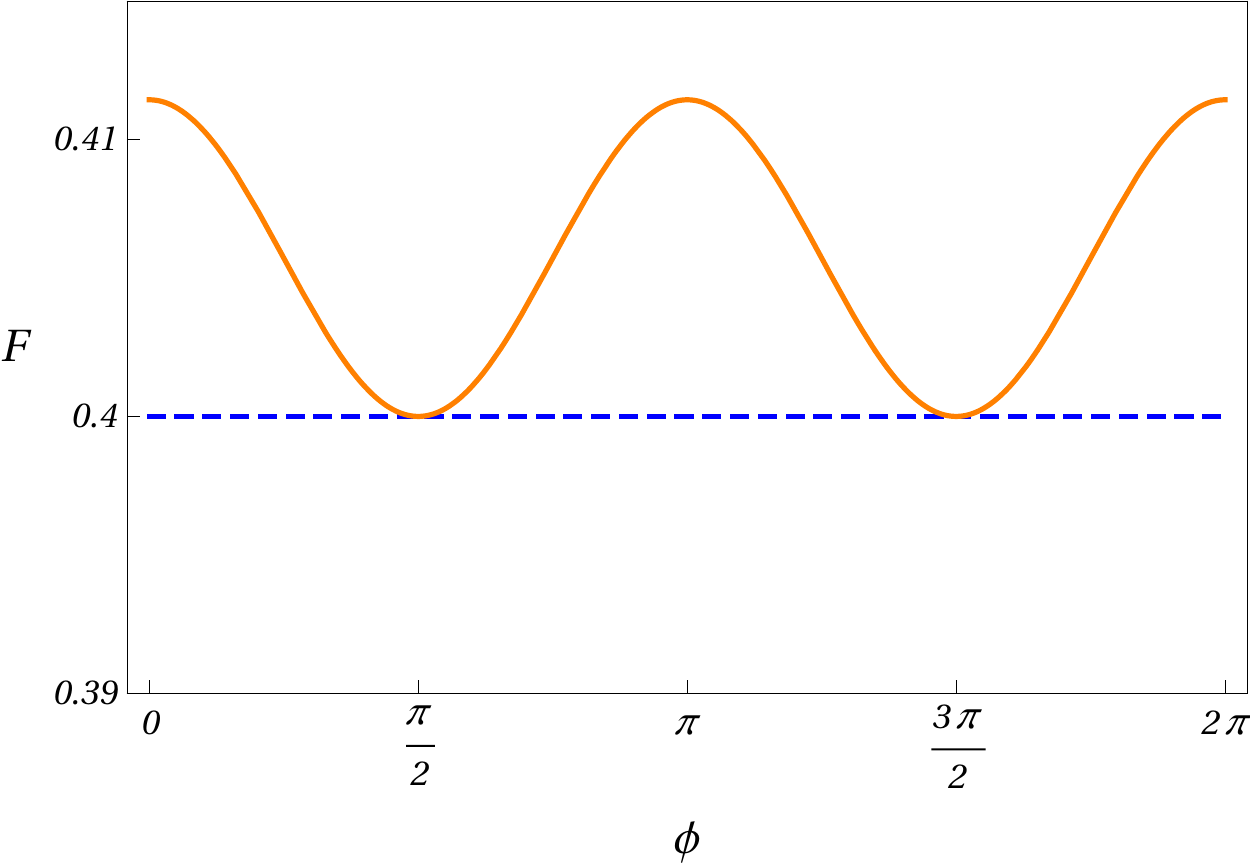}
 \caption{\it 
 Modulation of the response function $F$ of eq \eqref{detresf} as the interferometer rotates an angle $\phi$ 
around an axis aligned with one of the interferometer arms. We
assume that this arm points along the earth's rotation axis, and we choose a reference frame so that it
corresponds to the direction $\hat x$.  The dashed blue line shows the case with no anisotropy, $Q_{ij}=0$, whereas the orange line shows the case with one component $Q_{33}\,=\,-Q_{11}\,=\,0.1$ turned on. 
}
\label{fig:response}
\end{center} 
\end{figure}

    The total signal detected by an interferometer can decomposed as $S(t)\,=\,s(t)+n(t)$, with $n(t)$ the noise
    and $s(t)$ the contribution due to the gravitational wave. The relation between $s(t)$ and the mode $h_{ij}$
    can be written as \cite{Maggiore:1999vm}
    \be
    s(t)\,=\,{D}^{ij}\,h_{ij}(t)\,,
    \ee
  with ${D}_{ij}$ the detector tensor. For ground-based interferometers with arm directions $\hat u$ and $\hat v$ it reads
  \be
  {D}^{ij}\,=\,\frac12\,\left( 
\hat u^i \hat u^j-\hat v^i \hat v^j
  \right)\,. 
  \ee 
  We introduce the detector pattern function 
  \be
  F^{\lambda}({\bf k})\,=\,
  D^{ij}\,e^{(\lambda)}_{ij}( {\bf k})
\,. \ee 
The single detector response function, as defined for example in  Section 3 of \cite{Maggiore:1999vm}   is
the proportionality constant $F$ between   
the equal time 2pt function of the ground-based interferometer signal and the integral over frequencies of the amplitude of the primordial power spectrum ${\cal P}^{(0)}$.
 Using the techniques
  developed in \cite{Allen:1996gp}, we find that the response function reads in our case
 %
 %
%
  \be \label{detresf}
  F\,=\,\frac25
   -\frac{16}{35}\,{\cal Q}_{ij}{D}_{ik} D_{kj}\,.
  \ee
The first contribution $2/5$ 
 is the well known  
response function for an isotropic SGWB 
for a single, ground-based Michelson interferometer (see
 \cite{Maggiore:1999vm}, Section 3). The additional  contribution is  
instead new, and  contains the anisotropic contributions of \eqref{tens-anis} to the interferometer response function: notice
that it depends on the detector tensor ${D}_{ij}$.

 As pointed out in \cite{Allen:1996gp}, a response function
as eq \eqref{detresf} can lead to a diurnal modulation of a ground-based interferometer signal, since the interferometer
arms change orientation with respect to the anisotropy parameter ${\cal Q}_{ij}$  as the earth rotates around its axis.
 As a representative example, 
we plot in Fig \ref{fig:response} the modulation of the interferometer response function as the interferometer arms make a full rotation
around the axis of rotation of the earth, when some  of the components of the anisotropy parameter  ${\cal Q}_{ij}$ is turned on. From the figure we learn
that if we can probe percent variations of the interformeter response function  -- corresponding to diurnal modulations
of the same order in the stochastic background -- then we can probe anisotropies of the same magnitude: they 
 correspond to values of $f_{\rm NL}^{\rm sqz}$ of the order of one hundred, which can be achieved in the model
 we are considering (see Appendix \ref{app:concrete}).

It would be very interesting to study quantitatively whether current and future ground-based
interferometers  can 
 set constraints on the size of $Q_{ij}$ for realistic values of  $\hat f_{\rm NL}^{\rm sqz}$, by
 studying correlations among signals from different instruments in the presence of the
 primordial quadrupolar asymmetry,  and by computing
 the corresponding signal-to-noise ratio.
We plan to further develop these topics in a future publication. 

\section{Conclusions}

We investigated the consequences of a non-attractor phase of cosmological evolution
for the dynamics of primordial tensor modes, focussing on the properties of primordial
tensor non-Gaussianity in scenarios
with non-minimal couplings of gravity to the scalar sector. Thanks to a tensor duality, we have been able to analytically
compute the properties of the tensor bispectrum during this phase. 
We have shown that the  tensor bispectrum is 
 enhanced   in the squeezed limit 
with respect
to standard slow-roll scenarios, and can parametrically violate Maldacena's consistency relations.
 Moreover, tensor non-Gaussianity exhibits a  scale dependence  characteristic  of our set-up,
 that can help to distinguish our model from other scenarios with large tensor non-Gaussianity.  
 Squeezed tensor non-Gaussianity
induces a characteristic quadrupolar anisotropy on the power spectrum of 
 the stochastic  background of primordial tensor perturbations. To make contact with gravitational wave experiments, we discussed  the 
 response function of a ground based Michelson interferometer to a  gravitational wave background with such a feature. 

\smallskip

Much work is left for the future. It would be interesting to apply our approach to 
more general scenarios then the ones considered so far, including theories of Beyond Horndeski or DHOST 
\cite{Zumalacarregui:2013pma,Gleyzes:2014dya,Langlois:2015cwa,Crisostomi:2016czh,BenAchour:2016fzp}. This would
also allow one to study in more general terms the transition phase between attractor and non-attractor, and related possible 
instabilities associated with   violations of energy conditions (see the discussion in the Appendix
of \cite{Mylova:2018yap}). At the phenomenological
level, it would be important to further investigate to what extent gravitational wave experiments
can probe the quadrupolar anisotropy in the tensor power spectrum induced by
squeezed non-Gaussianity, by computing the corresponding signal-to-noise
ratio for actual experiments, and relating it to the size of non-Gaussian observables. We plan 
to return on these topics soon.

\acknowledgments

It is a pleasure to thank Alessandro Conigli and Angelo Ricciardone for discussions. 
We are 
 partially supported by STFC grant ST/P00055X/1.

\begin{appendix}


\section{Technical appendix:\\ computation of two-point and three-point functions
for tensor modes}\label{app:tech}

In this Appendix, we go through the technical details of the results we discuss in the main text. We start showing that
tensor duality allows us to obtain an analytical expression for tensor mode functions during the non-attractor era
dual to a slow-roll phase: this
is the basic ingredient we need for then analytically evaluating two point and three point functions in the
non-attractor regime. For definiteness, the set-up we have in mind to realise this scenario is the same
of Subsection \ref{sec:review-model}, first discussed in \cite{Mylova:2018yap}. In this case, 
\bea
{\cal F}_T&\propto& 1/a^6 \hskip1cm c_T^2\,=\,{\cal F}_T/{\cal G}_T\,\propto\,{\rm constant}\hskip1cm {\text{ during non-attractor phase}}\,,
\\
{\cal F}_T&\propto& {\rm constant} \hskip0.3cm c_T^2\,=\,{\cal F}_T/{\cal G}_T\,\propto\,{\rm constant}\hskip1cm  {\text{ during slow-roll phase}}\,.
\eea

\smallskip

We start by discussing the  computation of the quadratic case, leading to the tensor
power spectrum. 
To
 investigate the behavior of the tensor fluctuations, we  define the canonically normalized tensor fluctuation $v_{ij} = z_T h_{ij}$ to re-write the leading order action \eqref{actssp} as
\beq \label{actssp3}
S_{T}^{(2)}\,=\,\frac12\,\int \d y \,\d^3 x
\,\left[  \left( 
v'_{ij} \right)^2- \,\left(\vec \nabla v_{ij}  \right)^2 +\frac{z_T''}{z_T}\,v_{ij}^2\right]\,.
\eeq
Using the expression   \eqref{sec-der}, this leads to the following equation of motion in Fourier space 
\beq\label{CME}
{
v_{\bf k}'' + \left( k^2 -\fr{2}{y^2} \right) v_{\bf k} = 0,}
\eeq
where we have used the following expansion for the canonical tensor perturbation
\begin{eqnarray}\label{cve}
v_{ij}(y, \mathbf{k})=\sum_s e_{ij}^{(s)}(\mathbf{k}) \bigg[
v_{\mathbf{k}}(y)a_s(\mathbf{k})
+v^*_{-\mathbf{k}}(y)a_s^\dagger(-\mathbf{k})
\bigg].
\end{eqnarray}
Here, $e_{ij}^{(s)}$ is the polarization tensor with the helicity states $s=\pm$,
satisfying $e_{ii}^{(s)}(\mathbf{k})=0=k_je_{ij}^{(s)}(\mathbf{k})$.
Our   normalisation is
\begin{eqnarray}\label{pol1}
e_{ij}^{(s)}(\mathbf{k})e_{ij}^{*(s')}(\mathbf{k})=\delta_{ss'},
\end{eqnarray}
and we choose the phase so that the following relations hold:
\begin{eqnarray}\label{pol2}
e_{ij}^{*(s)}(\mathbf{k})=e_{ij}^{(-s)}(\mathbf{k})=e_{ij}^{(s)}(-\mathbf{k}).
\end{eqnarray}
The commutation relation for the creation and annihilation operators is
\begin{eqnarray}\label{com1}
[a_s(\mathbf{k}), a_{s'}^\dagger(\mathbf{k}')] =(2\pi)^3\delta_{ss'}\delta(\mathbf{k}
-\mathbf{k}').
\end{eqnarray}
The mode equation in \eqref{CME} implies that the power spectrum of the fluctuations $v_{\bf k}$ is scale invariant on large scales. In order to see this, we note the solution to the differential equation in \eqref{CME} that reduces to the adiabatic vacuum on small scales which reads 
\beq\label{solc}
{
v_{\bf k}(y) = \fr{1}{\sqrt{2k}}\left(1 - \fr{i}{ky}\right) e^{-iky}}, 
\eeq
so that $\mathcal{P}_h \sim k^3 |v_k|^2 \propto k^0$ in the limit $-ky \to 0$,  as anticipated. 
Using \eqref{solc} and the formulae for $z_T$, we write the solution to the original tensor mode function by using $h_{\bf k} \equiv v_{\bf k}/z_T$,
\beq\label{mfna}
h_{\bf k}(y) = \frac{i \sqrt{2} H}{\sqrt{c_T \mathcal{F}_T k^3}}~ (1 + i k y)~ e^{-iky},
\eeq
which essentially has the same form as the one in the slow-roll approximation (see eq  (11) of \cite{Gao:2012ib}),
although in this case ${\cal F}_T$ is strongly time dependent, as discussed in Section \ref{sec:review-model} of the main text.
 Indeed,  $\mathcal{F}_T$ is evolving rapidly as $\propto a^{-6}$ during the non-attractor regime contrary to the case in slow-roll where $\mathcal{F}_T\simeq$ constant, see eqs \eqref{fprowis}. 
 
 \subsection*{The two point function and the tensor power spectrum}
 Starting from the two point function of tensor modes in Fourier space,
 the power spectrum of tensor fluctuations can be defined by $\mathcal{ P}_h=(k^3/2\pi^2)\mathcal{ P}_{ij,ij}$ using the following expressions\footnote{Sometimes, it is also convenient to use the quantity $P_h(\bf{k})\,\equiv\, \frac{2 \pi^2 \,
 \mathcal{ P}_h  (\bf{k})}{k^3}$.} 
\begin{eqnarray}
\langle h_{ij}(\mathbf{k})h_{kl}(\mathbf{k}')\rangle
&=&(2\pi)^3\delta(\mathbf{k}+\mathbf{k}')\mathcal{ P}_{ij,kl}(\mathbf{k}),
\\
\mathcal{ P}_{ij,kl}(\mathbf{k})&=&|h_{\mathbf{k}}|^2\Pi_{ij,kl}(\mathbf{k}),
\end{eqnarray}
where
\begin{eqnarray}\label{fourind}
\Pi_{ij,kl}(\mathbf{k})=\sum_se_{ij}^{(s)}(\mathbf{k})e_{kl}^{*(s)}(\mathbf{k}).
\end{eqnarray}
We learn that on super horizon scales $-k y \to 0$, one  still gets a scale invariant power spectrum of primordial tensor fluctuations during the non-attractor phase. 
\begin{eqnarray}\label{Psg}
\mathcal{ P}_h = \lim_{-k y\to 0} \fr{k^3}{2\pi^2}2|h_{\bf k}|^2 = \fr{2}{\pi^2}\fr{H^2}{\mathcal{F}_T c_T}.
\end{eqnarray}
However, since the tensor kinetic term evolves as $\mathcal{F}_T \propto a^{-6}$, the amplitude of the power spectrum grows on super-horizon scales and therefore it has to be evaluated at the end of the non-attractor era. These result shows an agreement with what one would expect from the duality arguments discussed in \cite{Ozsoy:2018flq}, namely the power spectrum grows on superhorizon scales while it preserves a scale invariant form. 

 \subsection*{The three point function and the tensor bispectrum}

The tensor bispectrum can be computed by employing
the in-in formalism (see e.g. \cite{Weinberg:2005vy}), starting from the three point function in Fourier space
\begin{eqnarray}\label{inin}
\nn&&\langle h_{i_1j_1}(\mathbf{k}_1)
h_{i_2j_2}(\mathbf{k}_2)
h_{i_3j_3}(\mathbf{k}_3)
\rangle 
\\&&
=-i\int^t_{t_i}\d t'~\langle~ \left[
 h_{i_1j_1}(t, \mathbf{k}_1)
 h_{i_2j_2}(t, \mathbf{k}_2)
 h_{i_3j_3}(t, \mathbf{k}_3), H_{\rm int}(t')\right]~\rangle,
\end{eqnarray}
where $t_i$ is some early time when the perturbation is
well inside the sound horizon and the interaction Hamiltonian is given by $ H_{\rm int}(t) = -\int \d^3 x~ \mathcal{L}_T^{(3)} $. We find it convenient to work with the time coordinate $y$ in evaluating the three point function. We   express all the time dependent quantities with respect to the time when  the non-attractor era ends, $y_{\rm end}$. Using \eqref{mfna}, the mode functions therefore take the following form
\beq\label{mfnaa}
h_{\bf k}(y) = \frac{i \sqrt{2} H}{\sqrt{c_T \mathcal{F}_T^{(\rm end)} k^3}}~ (1 + i k y)~ e^{-iky} \left(\frac{y_{\rm end}}{y}\right)^{3},
\eeq
where $^{(0)}$ denotes the value of the quantity at the beginning of the non-attractor era. Using the expansions in 
\eqref{cve} (and noting $h_{ij} = v_{ij}/z_T$), we take the commutators in \eqref{inin} by taking into account the relations between the polarization tensors \eqref{pol1},\eqref{pol2} and \eqref{com1}.

 This procedure yields to the following expression for three point function,
\begin{eqnarray}\label{inin}
\nn&&\langle h_{i_1j_1}(\mathbf{k}_1)
h_{i_2j_2}(\mathbf{k}_2)
h_{i_3j_3}(\mathbf{k}_3)
\rangle 
\\&&
= (2\pi)^3 \delta({\bf k}_1+{\bf k}_2+{\bf k}_3)~\mathcal{T}^{(\alpha)}_{i_1 j_i i_2 j_2 i_3 j_3} \bigg( 2~{\rm Im}[~\mathcal{I}^{(\alpha)}(y)~] \bigg),
\end{eqnarray}
where the form of tensorial part $\mathcal{T}^{(\alpha)}$ and the function $\mathcal{I}^{(\alpha)}(y)$ depends on the interaction term under consideration in the third order action \eqref{cubic-action}, denoted by the superscript $\alpha\,=\,\left( {\rm GR}, {\rm new}\right)$. 

\smallskip
\noindent
{$\bullet$ \underline{\bf The `GR' contribution}}

\smallskip
\noindent
We begin our calculations by the first term in the interaction Hamiltonian $ H_{\rm int}(t) = -\int \d^3 x~ \mathcal{L}_T^{(3)} $, using the interaction Lagrangian \eqref{cubic-action} which we denote by $\alpha \to {\rm GR}$.
 In this case the tensorial part is given by
\begin{eqnarray}\label{TGR}
\nn\mathcal{T}_{i_1j_1i_2j_2i_3j_3}^{({\rm GR})} =
\bigg\{
\Pi_{i_1j_1,ik}(\mathbf{k}_1)\Pi_{i_2j_2,jl}(\mathbf{k}_2)
\bigg[
k_{3k}k_{3l}\Pi_{i_3j_3,ij}(\mathbf{k}_3)
&-&\frac{1}{2}k_{3i}k_{3k}\Pi_{i_3j_3,jl}(\mathbf{k}_3)
\bigg]\\&+&5 ~ {\rm perms}~{\rm of}~ 1, 2, 3
~\bigg\}.
\end{eqnarray}
where recall the definition of the four-index tensor $\Pi$ in eq \eqref{fourind}.
Since the tensor modes evolve outside the horizon during the non-attractor regime, we need to evaluate the function $\mathcal{I}$ (and hence the bispectrum) at the end of the non-attractor era beyond which we assume the mode functions (as well as time dependent quantities such as ${\mathcal F}_T$) freeze-out, since we return into a standard slow-roll phase.
 We are therefore interested in the following expression
\beq\label{IGR}
\mathcal{I}^{(\rm GR)} = h_{\bf k_1}(y_{\rm end})h_{\bf k_2}(y_{\rm end})h_{\bf k_3}(y_{\rm end}) \int_{-\infty}^{y_{\rm end}} \d y' ~\fr{a^2(y') {\mathcal F}_T(y')}{4~ c_T}~h^{*}_{\bf k_1}(y')h^{*}_{\bf k_2}(y')h^{*}_{\bf k_3}(y'),
\eeq
where we take $y_i \to -\infty$ to ensure all modes of interest are inside the horizon initially and $y_{\rm end}$ denotes the end of non-attractor era. Noting the behavior of the mode functions in \eqref{mfna} and of the tensor kinetic function ${\mathcal F}_T \propto y^{6}$ during the non-attractor era, the function $\mathcal{I}^{(\rm GR)}$ contains integrals of the following form
\beq\label{intC0}
\mathcal{C}^{(\alpha)}_p=\int_{\infty}^{x_{\rm end}} \d x'~(x')^{-p} ~e^{-i x'},
\eeq
where we defined $x' \equiv -K y'$ with $K = \sum_i k_i$ and $|{\bf k_i}| = k_i$. For the GR contribution we are focussing on, $p$ is a positive integer with the following possible values, $p =\{5,4,3,2\}$. The result of the integration can be expressed in terms of the incomplete Gamma functions,
\beq\label{intC}
\mathcal{C}^{(\alpha)}_p = -(-i)^{1-p}~ \Gamma(1-p, i x_{\rm end})
\eeq 
which has the following series expansion for negative integer  values of its first argument
\beq\label{GammaFe}
\Gamma(-m,z) = \fr{1}{m!}\left(\fr{e^{-z}}{z^m}\sum_{k=0}^{m-1}(-1)^k(m-k-1)!~z^k+(-1)^m~ \Gamma(0,z)\right).
\eeq
The results above are valid as far as we are away from the origin $x_{\rm end} =0$ but diverges as $x_{\rm end}\to 0$. In the following, we will express our results for small but non-zero value of $x_{\rm end}\equiv -K y_{\rm end} < 1$. Moreover, notice that
the incomplete Gamma $\Gamma(0,z)$ (also known as the exponential integral $E_1(z)$) admits   the following useful power series expansion in terms of elementary functions,
\beq\label{ga0}
\Gamma(0,z) = -\gamma_E-\ln(z)-\sum_{k=1}^{\infty}\fr{(-z)^{k}}{k ~k!}, 
\eeq
where $\gamma_E$ is the Euler-Mascheroni constant.
 Following the discussion above, we use \eqref{intC} with \eqref{GammaFe} and \eqref{ga0} to present our results at next to leading order in $-K y_{\rm end} < 1$. Keeping in mind the full expression in \eqref{IGR}, we thus have 
\begin{align}
\nn 2~{\rm Im}[\mathcal{I}^{(\rm GR)}]& =\fr{(2\pi)^4 \mathcal{P}_h^{(\rm end)^2}}{\Pi_i~ k_i^3}\\&~~\nn \times -\fr{K}{64}\Bigg[\left(1-\fr{3}{K^3}\sum_{i\neq j} k_i^2 k_j -\fr{6}{K^3}\Pi_i k_i\right)(-Ky_{\rm end})^2\\& ~~~~~~~~~~~~-\fr{\pi}{4}\left(1-\fr{4}{K^3}\sum_{i\neq j} k_i^2 k_j -\fr{4}{K^3}\Pi_i k_i\right)(-Ky_{\rm end})^3\Bigg],
\end{align}
where we made use of the expression for the power spectrum in \eqref{Psg} evaluated at the end of the non-attractor era. With these ingredients, we can express the tensor three point function associated with the `GR' contribution as
\begin{eqnarray}\label{bsp}
\langle h_{i_1j_1}(\mathbf{k}_1)
h_{i_2j_2}(\mathbf{k}_2)
h_{i_3j_3}(\mathbf{k}_3)
\rangle 
= (2\pi)^7 \delta({\bf k}_1+{\bf k}_2+{\bf k}_3)\fr{\mathcal{P}^{(\rm end)^2}_h}{\Pi_i~ k_i^3}~\mathcal{A}^{(\rm GR)}_{i_1 j_i i_2 j_2 i_3 j_3}, 
\end{eqnarray}
where we defined $\mathcal{A}^{(\rm GR)}_{i_1 j_i i_2 j_2 i_3 j_3} \equiv \mathcal{A^{(\rm GR)}}(k_1,k_2,k_3)~ \mathcal{T}^{(\rm GR)}_{i_1 j_i i_2 j_2 i_3 j_3} $ with  $\mathcal{T}^{(\rm GR)}$ given in eq \eqref{TGR}, 

\begin{align}\label{AGR}
\nn \mathcal{A}^{(\rm GR)}& = - \fr{K}{64}\Bigg[\left(1-\fr{3}{K^3}\sum_{i\neq j} k_i^2 k_j -\fr{6}{K^3}\Pi_i k_i\right) (-Ky_{\rm end})^2\\& ~~~~~~~~~~~-\fr{\pi}{4}\left(1-\fr{4}{K^3}\sum_{i\neq j} k_i^2 k_j -\fr{4}{K^3}\Pi_i k_i\right) (-Ky_{\rm end})^3+\dots\Bigg],
\end{align}
and dots indicates sub-leading terms higher order in $-Ky_{\rm end}$. The amplitude $\mathcal{A}^{(\rm GR)}$ shows the scale dependence of the non-Gaussianity parametrized by the powers of $-K y_{\rm end} <1$  during the non-attractor era.

\smallskip
\noindent
{$\bullet$ \underline{\bf The `new' contribution}}

\smallskip
\noindent
In a similar fashion, we now move on to calculate the contribution to the tensor three point
function from the second term in the interaction Hamiltonian (see \eg \eqref{cubic-action}). Following the same steps as we did for the previous term, we write the three point
function in the same form as in \eqref{inin} with
\begin{eqnarray}\label{TNEW}
\mathcal{T}_{i_1j_1i_2j_2i_3j_3}^{({\rm new})} =
\Pi_{i_1j_1,ij}(\mathbf{k}_1)\Pi_{i_2j_2,jk}(\mathbf{k}_2)
\Pi_{i_3j_3,kl}(\mathbf{k}_3)
\end{eqnarray}
and
\beq\label{Inew}
\mathcal{I}^{(\rm new)} = -h_{\bf k_1}(y_{\rm end})h_{\bf k_2}(y_{\rm end})h_{\bf k_3}(y_{\rm end}) \int_{-\infty}^{y_{\rm end}} \d y' ~\fr{c_T^2 a\dot{\phi} X G_{5X}}{2}~h^{*'}_{\bf k_1}(y')h^{*'}_{\bf k_2}(y')h^{*'}_{\bf k_3}(y').
\eeq
Here a prime on the mode functions denotes a time derivative with respect to their argument $y'$. We note that since the integral contains these time derivatives, the calculation of the 3-pt function associated with  this contribution is  more involved compared to the previous case. In order to deal with the integral in \eqref{Inew}, we note the time derivative of the complex conjugated mode function as 
\beq\label{mfnacp}
h^{*'}_{\bf k}(y) = -\frac{i \sqrt{2} H}{\sqrt{c_T \mathcal{F}_T^{(\rm end)} k^3}} \left(\frac{y_{\rm end}}{y}\right)^{3}~ \left\{-\fr{3}{y}(1 - i k y)+k^2 y\right\}~ e^{iky},
\eeq
which reflects the rapid change of the mode functions after horizon crossing ($-ky \to 0$). Noting the time dependence of the functions inside the integral, \ie $\dot{\phi} \propto a^{-3}$, $X \propto a^{-6}$ and $G_{5X} \propto a^{3}$, we see that we need to deal with integrals that has the same form as in \eqref{intC0} with $p=\{7,6,5,4,3,2,1\}$.
Therefore, repeating the same steps as we did for the ``GR'' term, we obtain the following results for $2 {\rm Im}[\mathcal{I}^{(\rm new)}]$ at next to leading order in $-K y_{\rm end}$, 
\begin{align}
\nn 2~{\rm Im}[\mathcal{I}^{(\rm new)}] =-\fr{(2\pi)^4 \mathcal{P}_h^{(\rm end)^2}}{\Pi_i~ k_i^3}~\fr{3H\mu^{(\rm end)}}{16\mathcal{G}^{(\rm end)}_T} &\Bigg\{\left(K^3-3\sum_{i\neq j} k_i^2 k_j -6~\Pi_i k_i\right)\\&-\fr{1}{4}\Bigg(K^3-5\sum_{i\neq j} k_i^2 k_j +\fr{2}{K^2}\sum_{i\neq j}k_i^3k_j^2\Bigg)(-Ky_{\rm end})^2\Bigg\},
\end{align}
where we defined $\mu^{(\rm end)} \equiv\dot{\phi}^{(\rm end)}X^{(\rm end)}G_{5X}^{(\rm end)}$.
The contribution of this term to the 3pt function can be written  similarly to the expression in \eqref{bsp} where we define the amplitude $\mathcal{A}^{(\rm new)}_{i_1 j_i i_2 j_2 i_3 j_3} \equiv \mathcal{A^{(\rm new)}}(k_1,k_2,k_3)~ \mathcal{T}^{(\rm new)}_{i_1 j_i i_2 j_2 i_3 j_3} $ and 
\begin{align}\label{ANEW}
\nn\mathcal{A^{(\rm new)}}= -\fr{3H\mu^{(\rm end)}}{16\mathcal{G}^{(\rm end)}_T} &\Bigg\{\left(K^3-3\sum_{i\neq j} k_i^2 k_j -6~\Pi_i k_i\right)\\&-\fr{1}{4}\Bigg(K^3-5\sum_{i\neq j} k_i^2 k_j +\fr{2}{K^2}\sum_{i\neq j}k_i^3k_j^2\Bigg)(-Ky_{\rm end})^2\Bigg\} .
\end{align}
This result shows that the contribution of the new term to the amplitude of the bispectrum has a scale independent piece plus a scale dependent subleading term, which becomes small as $-Ky_{\rm end}\to 0$. The difference between the scale dependence of the $\mathcal{A^{(\rm new)}}$ and $\mathcal{A^{(\rm GR)}}$ can be understood by analyzing the time dependence of each term in the interaction Lagrangian \eqref{cubic-action}. For example, during the non-attractor phase the new term can be written schematically as $a^{-6} h h h$ whereas the GR term scales with scale factor as $a^{-8} h h h$ where we have suppressed the indices on the metric. This explains why the contribution from each term differs by a factor $y_{\rm end}^{-2}$ at leading order in the amplitude of the bispectrum.

\section{Disformal transformation and tensor non-Gaussianity}\label{app:disf}
The general quadratic action for tensors  in \eqref{actssp} can be transformed into a form identical to the action for tensor fluctuations in general relativity (GR) by applying a disformal and conformal transformation to the metric successively \cite{Creminelli:2014wna,Baumann:2015xxa}. In this appendix, we discuss the implications\footnote{See also \cite{Domenech:2015hka} for a general analysis of the consequences of disformal transformations on cosmological fluctuations.} of such transformations for the background dynamics and for the tensor bispectrum we discussed earlier in Section \ref{sec:nonG} and Appendix \ref{app:tech}. For our system, the corresponding combination of disformal and conformal transformation is given by
\beq
g_{\mu\nu} \to \frac{\mathcal{F}_T}{c_T} \left[g_{\mu\nu}+(1-c_T^2)n_\mu n_\nu\right],
\eeq
which corresponds to the following re-labeling of the time coordinate and re-definition of the scale factor,
\beq\label{DandC}
\d \hat{t} = (c_T \mathcal{F}_T)^{1/2} \d t, ~~~~ \hat{a}(\hat{t}) = \left(\frac{\mathcal{F}_T}{c_T}\right)^{1/2}a(t).
\eeq
Using the transformations in \eqref{DandC}, the quadratic action in \eqref{actssp} take the standard form that appear in GR,
\bea\label{EfST}
\nn S_T^{(2)} &=& \frac{1}{8}\int \d \hat{t}~ \d^3 x~ \hat{a}(\hat{t})^3 \left[(\partial_{\hat{t}} h_{ij})^2 - \frac{1}{\hat{a}(\hat{t})^{2}}(\vec{\nabla}h_{ij})^2\right],\\
&=& \frac{1}{2}\int \d y~ \d^3 x~ \frac{\hat{a}^2}{4} \left[(\partial_{y} h_{ij})^2 - (\vec{\nabla}h_{ij})^2\right],
\eea
where in the second line we have used the fact that the conformal time in the GR frame is defined by the coordinate $y$ , namely $\d \hat{t}/ \hat{a}(\hat{t})\equiv \d y$, which can be seen by combining the expressions given in \eqref{DandC}.

In order to describe the time evolution of the background in the Einstein Frame, we make use of the relation between two scale factors in \eqref{DandC} together with the fact that $\mathcal{F}_T \propto a^{-6}$ and $a \propto -1/y$. This procedure leads to the conclusion that, universe appear to be collapsing as in a dust dominated universe, that is
\beq\label{Efsf}
\hat{a} \propto y^2,
\eeq
as $y \to 0$. Similarly, we can relate the Hubble rate in the Einstein frame, $\hat{H}=\d \ln \hat{a}/\d \hat{t}$, to the Hubble rate in the Jordan frame using \eqref{DandC}, which leads to 
\beq\label{hH}
\hat{H} = -\frac{2 H}{(c_T\mathcal{F}_T)^{1/2}}  \propto \hat{a}^{-3/2},
\eeq
as expected from a dust dominated universe. Using the transformation \eqref{hH} for the Hubble rate, the power spectrum of tensor fluctuations in the Jordan frame can be expressed in terms of the quantities in the Einstein frame as
\beq
\mathcal{P}_h = \frac{\hat{H}^2}{2\pi^2}  \propto \hat{a}^{-3}.
\eeq
This expression reflects the equivalence of the interpretation of the results in both frames. In the Einstein frame, the power spectrum of tensor fluctuations also appear to be increasing during the transient   collapsing\footnote{Note that similar to the time span  $y_0 < y < y_{\rm end}$ of the non-attractor era in the Jordan frame, the collapsing phase in the Einstein frame will last for a finite time.} phase as $\hat{a} \to 0$.

The equivalence of the results in both frames also extends to the observables such as the tensor non-Gaussianity. In the following, we prove that the calcuation of the bispectrum is equivalent in both frames. For this purpose, we first realize from \eqref{EfST} and \eqref{Efsf} that the canonical variable $v_{ij} = z_T h_{ij}$ with $z_T = \hat{a}/2$ satisfies the same equation in Fourier space similar to the case during the non-attractor phase (See, \eg eq. \eqref{CME}). Therefore, in the Einstein frame, the mode functions that reduces to the Bunch-Davies vacuum is given by,
\beq
h_{\bf k} = \frac{-i\hat{H}}{\sqrt{2k^3}}\left(1+ i k y \right) e^{-i k y}.
\eeq
Notice that using the relation \eqref{hH}, the mode functions appear to have the same form as the one in the non-attractor phase \eqref{mfnaa}:
\beq\label{mf}
h_{\bf k}=\frac{i\sqrt{2}H}{\sqrt{c_T\mathcal{F}_T k^3}}\left(1+ i k y \right) e^{-i k y},
\eeq
where $\mathcal{F}_T = \mathcal{F}_T^{(\rm end)}(y/y_{\rm end})^6$. In order to establish the equivalence of the in-in calculation in both frames, we therefore only need to focus on the time dependence of the interaction Hamiltonian in the Einstein frame, which is given by
\beq
H_{\rm int}(y) = -\int \d^3 x~ \bigg[\frac{Q_{\rm new}(y)}{12}~h'_{ij}h'_{jk}h'_{ki}+ \frac{\hat{a}^2(y)}{4} \left(h_{ik}h_{jl}
-\frac{1}{2}h_{ij}h_{kl}\right)\partial_k\partial_lh_{ij}\bigg],
\eeq
where prime denotes a time derivative w.r.t $y$ and we have defined the time dependent pre-factor of the new interaction as
\beq
Q_{\rm new} = \frac{\hat{a}\mathcal{F}_T^{3/4}}{\mathcal{G}_T^{5/4}} X\dot{\phi}G_{5X}.
\eeq
We proceed the in-in calculation in the Einstein frame by defining analogues of the functions $\mathcal{I}^{(\rm new)}$ and $\mathcal{I}^{(\rm GR)}$ that we defined earlier in the Jordan frame. Following the same steps as we before, these functions in the Einstein frame is given by
\begin{align}
\nn \hat{\mathcal{I}}^{(\rm GR)} &= h_{\bf k_1}(y_{\rm end})h_{\bf k_2}(y_{\rm end})h_{\bf k_3}(y_{\rm end}) \int_{-\infty}^{y_{\rm end}} \d y' ~\fr{\hat{a}^2(y')}{4}~h^{*}_{\bf k_1}(y')h^{*}_{\bf k_2}(y')h^{*}_{\bf k_3}(y'),\\
\hat{\mathcal{I}}^{(\rm new)}& = -h_{\bf k_1}(y_{\rm end})h_{\bf k_2}(y_{\rm end})h_{\bf k_3}(y_{\rm end}) \int_{-\infty}^{y_{\rm end}} \d y' ~\frac{Q_{\rm new}(y')}{2}~h^{*'}_{\bf k_1}(y')h^{*'}_{\bf k_2}(y')h^{*'}_{\bf k_3}(y'),
\end{align}
where $y_{\rm end}$ denotes the end of the collapsing phase. Noting $\hat{a} \propto y^2$
 and $Q_{\rm new}\propto \hat{a}^{5/2}\propto y^5$ and the mode functions in \eqref{mf}, we see that we need to deal with identical integrals in the calculation of bispectrum amplitude in the Einstein frame. In particular, defining the dimensionless variable $x' = -K y'$, integrals have the same form as before (See for example, eq. \eqref{intC0}):
 \beq
 \mathcal{C}^{(\alpha)}_p=\int_{\infty}^{x_{\rm end}} \d x'~(x')^{-p} ~e^{-i x'},
 \eeq
where $\alpha$ labels the new or the GR term with $p=\{7,6,5,4,3,2,1\}$ and $p=\{5,4,3,2\}$. Therefore, as expected, one can reach at the same results we derived earlier for the amplitude of the bispectrum in the Einstein frame.

\section{An explicit scenario with large tensor non-Gaussianity}\label{app:concrete}





\bigskip

In this appendix we apply the results presented in the main text, for the amplitude of tensor non-Gaussianity during a phase of non-attractor inflation,  to a concrete model.  We use the model introduced in Section \ref{sec:review-model}, which is based on the Horndeski theory of gravity,  choosing the Horndeski functions as in \eqref{GF}.  The model is discussed in detail in \cite{Mylova:2018yap}, and we present in this appendix only a brief summary of the relevant dynamics and parameter space.  In particular, our purpose here is to identify a consistent parameter space for the model, which gives rise to a large tensor bispectrum during a non-attractor phase.

  We allow the free parameters, $\rho, \alpha, \nu, \delta, \beta$ and $\sigma$ in \eqref{GF}  to take different values during three different phases of inflationary evolution.  Provided that the parameters satisfy certain relations, the equations of motion admit a solution with a constant Hubble parameter and continuous metric and $\dot{\phi}$, consisting of two slow roll inflationary phases, during which $\dot{\phi}=const$, connected by a transient non-attractor self-accelerating de Sitter phase that is tensor dual to the initial slow roll phase, with $\dot{\phi} \propto 1/a^3$.   As we show in \cite{Mylova:2018yap}, such a solution is possible provided that the parameters satisfy $\rho=2\delta+3\beta+4\sigma$ and $\nu=0$, during the non-attractor phase.

Whilst the non-minimal derivative couplings between metric and scalar in Horndeski Lagrangians have been chosen to allow a non-attractor inflationary phase, tensor dual to slow roll, care must be taken so that they do not also introduce ghost or gradient instabilities in the tensor and scalar fluctuations.  As discussed in detail in \cite{Mylova:2018yap}, the stability constraints restrict the parameter space to $\rho<0$, $\beta>0$, $\sigma<0$, $\delta<0$, $\beta+9 \sigma<0$, $\beta+3 \sigma >0$ and $f_{s}>0$, during the non-attractor era, where 
\beq
f_s=\frac{-2\beta^2+\delta(\beta+9\sigma)+3\sigma(\beta+3\sigma)}{6(3\beta+\delta+6\sigma)}.
\eeq


Referring to \cite{Mylova:2018yap} for the explicit solution for $\dot{\phi}$ in the non-attractor phase, we now write the non-linearity parameter given in \eqref{fnlg} in terms of the model parameters:
\beq\label{hmu}
\hat{f}^{+++}_{\rm NL(\rm new)}=\frac{135}{512}\frac{H\mu^{\rm end}}{\mathcal{G}_T^{\rm end}} = -\frac{135}{512}\frac{3\sigma}{2(\beta + 3\sigma)}.
\eeq  
It is clear from the expression above that bispectrum amplitude can be large, in the limit $\beta \to -3\sigma$. In order to parametrize the proximity to this limit and hence the enhancement of the bispectrum, we thus set $\beta = -3\sigma + \epsilon$ where $0<\epsilon\ll -3\sigma$. Notice that this parametrization guarantees that the stability conditions $\beta + 3\sigma >0$ and $\beta + 9\sigma < 0$ are satisfied, when $\sigma<0$. The condition $f_s>0$ can then be satisfied by fixing the parameter $\delta$ within the range:
\beq\label{d2}
3\sigma - 3\epsilon + \frac{5\epsilon^2}{6\sigma}<\delta < 3\sigma - 3\epsilon. 
\eeq
We can then set the final parameter $\rho$ using $\rho =2\delta	+3\beta+4\sigma$ as required 
by the equations of motion in the non-attractor self-accelerating de Sitter background.

The expressions \eqref{hmu} and \eqref{d2} imply that in order to enhance the bisepctrum by an amount $\epsilon^{-1}$, we require a cancellation between $\beta$ and $3|\sigma|$ at the order of $\epsilon$, together with a fine-tuning of $\delta$ (and thus $\rho$) at the order of $\epsilon^2$.  As shown in \cite{Mylova:2018yap}, an enhancement for the tensor power spectrum is also achieved, by choosing the parameter $\alpha$ to be suppressed in the non-attractor regime, with respect to the slow-roll regime preceding it.  The scalar power spectrum is also enhanced, by a smaller factor.  For example, for the parameters considered in \cite{Mylova:2018yap} -- $\sigma=-0.2$, $\beta=1.5$, $\delta = -4$, $\alpha_{\rm sr}/\alpha_{\rm na} = 10^{-3}$ -- the scalar power spectrum is enhanced by a factor $5.3 \times 10^6$, the tensor power spectrum by a factor of $5.0\times 10^7$ and the non-linearity parameter, \eqref{hmu}, is $\hat{f}^{+++}_{\rm NL(\rm new)}=0.088$.  Fine-tuning instead $\sigma=-1$, $\beta=3.001$, $\delta=-3.003007$ and $\alpha_{\rm sr}/\alpha_{\rm na} = 10^{-6}$, the scalar power spectrum is enhanced by a factor $5.5 \times 10^7$, the tensor power spectrum by a factor of  $1.3\times 10^8$ and the non-linearity parameter, \eqref{hmu}, is $\hat{f}^{+++}_{\rm NL(\rm new)}=395.5$.   We see that -- with sufficient fine-tuning -- large, potentially observable tensor non-Gaussianities can occur.


\end{appendix}
\addcontentsline{toc}{section}{References}
\bibliographystyle{utphys}

\bibliography{refs}

\end{document}